\documentclass[twocolumn,showpacs,preprintnumbers,amsmath,amssymb,nofootinbib]{revtex4}

\usepackage{graphicx}
\newcommand{\bff}[1]{{\mbox{\boldmath $#1$}}}

\begin{document}
\title{Time-odd mean fields in covariant density functional theory: Rotating systems.}
\author{A.\ V.\ Afanasjev and H.\ Abusara}
\affiliation{Department of Physics and Astronomy, Mississippi State 
University, Mississippi State, Mississippi 39762, USA}
\date{\today}

\begin{abstract}

Time-odd mean fields (nuclear magnetism) and their impact on physical observables in 
rotating nuclei are studied in the framework of covariant density functional theory (CDFT). 
It is shown that they have profound effect on the dynamic and kinematic moments of inertia. 
Particle number, configuration and rotational frequency dependences of their impact on the 
moments of inertia have been analysed in a systematic way. 
Nuclear magnetism can also considerably modify the band crossing features such as crossing 
frequencies and the properties of the kinematic and dynamic moments of inertia in the band crossing 
region. The impact of time-odd mean fields on the moments of inertia in the regions away from band 
crossing only weakly depends on the relativistic mean field parametrization, reflecting good 
localization of the properties of time-odd mean fields in CDFT. The moments of inertia of 
normal-deformed nuclei considerably deviate from the rigid body value. On the contrary, 
superdeformed and hyperdeformed nuclei have the moments of inertia which are close to rigid 
body value. The structure of the currents in rotating frame, their microscopic origin and
the relations to the moments of inertia have been systematically analysed.
The phenomenon of signature separation in odd-odd nuclei, induced by time-odd mean fields, 
has been analysed in detail.
\end{abstract}

\maketitle

%%%%%%%%%%%%%%%%%%%%%%%%%%%%%%%%%%%%%%%%%%%%%%%%%%%%%%%%%%%%%%%%%%%
\section{Introduction}
\label{Sect-intro}
%%%%%%%%%%%%%%%%%%%%%%%%%%%%%%%%%%%%%%%%%%%%%%%%%%%%%%%%%%%%%%%%%%%

   The development of self-consistent many-body theories aiming at the description 
of low-energy nuclear phenomena provides the necessary theoretical tools for an 
exploration of the nuclear chart into known and unknown regions. Theoretical 
methods (both relativistic \cite{VALR.05} and non-relativistic \cite{BHR.03}) formulated 
within the framework of density functional theory (DFT) are the most promising 
tools for the global investigation of the properties of atomic nuclei. The power 
of the DFT models is essentially unchallenged in medium and heavy mass nuclei 
where 'ab-initio' type few-body calculations are computationally impossible and 
the applicability of the spherical shell model is restricted to a few regions in 
the vicinity of doubly shell closures.

  The {\it mean field} is a basic concept of every DFT. One can specify  {\it time-even} 
and {\it time-odd} mean fields \cite{DD.95,AR.00} dependent on the response of these fields 
to the action of time-reversal operator. While the properties of time-even mean fields in 
nuclear density  functionals are reasonably well understood and defined \cite{BHR.03,VALR.05}, 
there are still many unknowns in our knowledge of time-odd mean fields which appear only in 
the nuclear systems with broken time-reversal symmetry. This is especially 
true in the covariant density functional theory (CDFT) \cite{VALR.05} where only few articles
were dedicated to the study of time-odd mean fields (see Ref.\ \cite{AA.10} for review). 
Note that the effects, produced by the magnetic potential in the Dirac equation and called 
as \textit{nuclear magnetism } (NM) \cite{KR.89} in the framework of the CDFT, are due to 
time-odd mean fields.

  Rotating nuclei represent a system which is strongly affected by time-odd mean 
fields. The representative studies of few examples \cite{PWM.85,KR.93,DD.95,AKR.96,AR.00} 
clearly show that the kinematic and dynamic moments of inertia of the nuclei rotating in 
collective manner are considerably affected by time-odd mean fields.  It was shown in the 
CDFT framework \cite{AR.00} that microscopic mechanism of this modification is traced back 
to the modifications of the expectation values  of the single-particle angular 
momentum $\langle \hat{j}_x \rangle_i$ in the presence of NM.  The contribution to 
$\langle \hat{j}_x \rangle_i$ due to NM is defined as
\begin{eqnarray}
\Delta \langle j_x \rangle_i= \langle \hat{j}_x \rangle^{NM}_i
- \langle \hat{j}_x \rangle^{WNM}_i
\label{Deltaj}
\end{eqnarray}
where the subscripts NM and WNM indicate the values obtained in the calculations 
with and without NM, respectively. The $\Delta \langle j_x \rangle_i$ is positive at the 
bottom and negative at the top of the $N$-shell \cite{AR.00}. The absolute value of 
$\Delta \langle j_x \rangle_i$ correlates with the absolute value of $\langle \hat{j}_x \rangle_i$. 
Note that the contributions to 
$\langle \hat{j}_x \rangle_i$ due to NM are small in the middle of the shell. The
$\Delta \langle j_x \rangle_i$ contributions can be decomposed into the contributions due
to spin ($\Delta \langle s_x \rangle_i$) and orbital ($\Delta \langle l_x \rangle_i$) angular 
momenta, which have complicated dependences both on the frequency and the structure of the 
single-particle orbital under study \cite{AR.00}. Similar features are expected also in 
non-relativistic DFT \cite{AR.00}.

  The changes in the alignment properties of the single-particle orbitals induced by NM 
(Eq.\ (\ref{Deltaj})) reflect themselves also in physical observables such as effective 
alignments and the energy splittings between signature partner orbitals (signature splitting), 
measured experimentally \cite{AR.00}. Moments of inertia and effective alignments in normal- 
and superdeformed nuclei in different parts of nuclear chart 
\cite{KR.93,AKR.96,ALR.98,ARR.99,AF.05,CRHB,VALR.05,A.08} are well described by the 
parametrizations which include non-linear self-couplings only for the $\sigma$-meson (see 
Table 1 in Ref.\ \cite{AA.10}). This fact strongly suggests that NM is well accounted in this 
type of the relativistic mean field (RMF) parametrizations. In addition, NM can have an impact 
on the energy gap between the yrast and excited configurations in local minima (as illustrated 
on the example of the hyperdeformed minima in Ref.\ \cite{AA.09}), on the terminating states 
\cite{A.08} and on the additivity of angular momentum alignments \cite{MADLN.07}.

  A systematic investigation of time-odd mean fields in one- (two-) particle states in 
odd (odd-odd) non-rotating nuclei has been performed in our previous article \cite{AA.10}. 
The current manuscript is a continuation of our efforts aimed on comprehensive understanding of 
time-odd mean fields in CDFT. Its goal is a systematic study of time-odd mean fields and 
their manifestation in rotating nuclei.

   Table 1 in Ref.\ \cite{AA.10} shows large variety of the parametrizations of the 
RMF Lagrangian. The investigation of all these 
parametrizations is definitely beyond the scope of this study. Thus, the present 
investigation has been focused on the study of time-odd mean fields in the CDFT with 
the parametrizations of the RMF Lagrangian including only non-linear self-couplings of 
the $\sigma$-meson (the group A of the parametrizations in 
Table 1  of Ref.\ \cite{AA.10}). So far, only this group of parametrizations has been used 
in the study of rotating nuclei \cite{AKR.96,ALR.98,ARR.99,AF.05,CRHB,A250,VALR.05,AA.08,A.08}. 
The results of the study of time-odd mean fields in the groups B, C, and D of the parametrizations 
of meson-coupling models (Table 1 in Ref.\ \cite{AA.08}) as well as within the point-coupling 
models will be presented in a forthcoming manuscript.

  The manuscript is organized as follows. The cranked RMF theory and its details related to 
time-odd mean fields in rotating nuclei are discussed in Sec.\ \ref{theory}. Section 
\ref{Sect-band-crossing} is devoted to the analysis of the impact of time-odd mean fields on 
band crossing features.  Particle number and deformation dependences of the impact of NM on 
the moments of inertia are considered in Sec.\ \ref{pndefdep-mom}. Currents (and their 
single-particle origin) in the intrinsic frame of rotating nuclei are discussed in Sec.\ 
\ref{Sec-currents}. Frequency and configuration dependences of the impact of NM on the moments 
of inertia are analyzed in Sec.\ \ref{Freq-dep}. Parametrization dependence of the NM contributions 
to the moments of  inertia are discussed in Sec.\ \ref{Sec-par-dep}. Sec.\ \ref{Sec-term-states} 
is devoted to the study of time-odd mean fields in terminating states.  The phenomenon of 
signature separation in odd-odd nuclei is investigated in Sec.\ \ref{Sec-sign-sep}. 
Finally, Sec.\ \ref{Sec-final} 
reports the main conclusions of our work.

%%%%%%%%%%%%%%%%%%%%%%%%%%%%%%%%%%%%%%%%%%%%%%%%%%%%%%%%%%%%%%%%%%%%%%%%%%%
\section{Theoretical formalism}
\label{theory}
%%%%%%%%%%%%%%%%%%%%%%%%%%%%%%%%%%%%%%%%%%%%%%%%%%%%%%%%%%%%%%%%%%%%%%%%%%%

 The results presented in the current manuscript have been obtained 
mainly in the framework of Cranked Relativistic Mean Field (CRMF) theory 
\cite{KR.89,KR.93,AKR.96}. Only in a few cases the results obtained within 
the Cranked Relativistic Hartree-Bogoliubov (CRHB) theory \cite{CRHB} are 
shown. The CRMF theory has been successfully employed for the description of 
rotating nuclei (see Refs.\ \cite{VALR.05,AA.08} and references therein) in 
which time-odd mean fields play an important role. In this theory the pairing 
correlations are neglected which allows to better isolate the effects induced 
by time-odd mean fields. The most important features of the CRMF formalism 
related to time-odd mean fields are outlined below (for more details see 
Refs.\ \cite{KR.89,AKR.96}), while the details of the CRHB theory are presented 
in Ref.\ \cite{CRHB}.

In the Hartree approximation, the stationary Dirac equation for the nucleons in the 
rotating frame (in one-dimensional cranking approximation) is given by
\begin{eqnarray}
(\hat{h}_{D}-\Omega_x  \hat{J}_x) \psi_i=\varepsilon_i \psi_i
\end{eqnarray}
where $\hat{h}_{D}$ is the Dirac Hamiltonian for the nucleon with mass $m$
\begin{eqnarray}
\hat{h}_{D}={\bff{\alpha}}(-i{\bff{\nabla}}-{\bff{V}}({\bff r}))~+~V_{0}({\bff r})~
+~\beta (m+S({\bff r}))
\end{eqnarray}%
and the term 
\begin{eqnarray}
-\Omega_x  \hat{J}_x = -\Omega_x \left( \hat{L}_x+\frac{1}{2}\hat{\Sigma}_x \right)   
\end{eqnarray}
is just the Coriolis term. Note that the rotational frequency $\Omega_x$ along the 
$x$-axis is defined from the condition that the expectation value of the total angular
momentum at spin $I$ has a definite value \cite{Ing.54}
\begin{eqnarray}
J(\Omega_x)=\langle {\Phi }_{\Omega }\mid \hat{J}_{x}\mid {\Phi
}_{\Omega }\rangle=\sqrt{I(I+1)}.  \label{cranking-condition}
\end{eqnarray}
The Dirac Hamiltonian contains the average fields determined by the mesons, i.e. the attractive 
scalar field $S({\bf r})$
\begin{eqnarray}
S({\bff r})=g_{\sigma} \sigma ({\bff r}),
\end{eqnarray}
and the repulsive time-like component of the vector field $V_{0}({\bff r})$
\begin{eqnarray}
V_0({\bff r}) = g_{\omega} \omega_0({\bff r}) + g_{\rho} \tau_3 \rho_0 ({\bff r}) + e 
\frac{1-\tau_3}{2} A_0 ({\bff r}).
\end{eqnarray}  
A magnetic potential ${\bff V} ({\bff r})$ 
\begin{eqnarray}
\bff{V}({\bff r})=g_{\omega }{\bff{\omega}}({\bff r})+ g_{\rho}\tau_3{\bff{\rho}}
({\bff r})+e\frac{1-\tau _{3}}{2}\bff{A}({\bf r})
\label{magnetic}
\end{eqnarray}
originates from the space-like components of the vector mesons. Note that in these
equations, the four-vector components of the vector fields $\omega^{\mu}$, 
$\rho^{\mu}$, and $A^{\mu}$ are separated into the time-like ($\omega_0$, $\rho_0$
and $A_0$) and space-like [${\bff \omega}=(\omega^x, \omega^y, \omega^z)$, 
${\bff \rho}=(\rho^x, \rho^y, \rho^z)$, and ${\bff A}=(A^x, A^y, A^z)$] components. 
In the Dirac equation the magnetic potential has the structure of a magnetic field.  
% Therefore the effect produced by it is called \textit{nuclear magnetism } (NM) 
% \cite{KR.89}. 

  The corresponding meson fields and the electromagnetic potential are determined by 
the Klein-Gordon equations
\begin{eqnarray}
\left\{ -\Delta
%-(\bm{\Omega\hat{L}})^{2}
+m_{\sigma}^{2}\right
\} ~\sigma({\bff r}) & = & -g_{\sigma} [\rho^n_s({\bff r}) + \rho^p_s ({\bff r})]
\nonumber \\
& & -g_{2}\sigma^{2}({\bff r})-g_{3}\sigma^{3}({\bff r}), \label{KGsigma} \\
\left\{ -\Delta
%-(\bm{\Omega\hat{L}})^{2}
+m_{\omega}^{2}\right\}
\omega_{0}({\bff r}) & = &g_{\omega} [\rho_v^n({\bff r}) + \rho_v^p({\bff r})],  
 \label{KGomega0} \\
\left\{ -\Delta
%-(\bm{\Omega\hat{J}})^2
+m_{\omega }^{2}\right\}
~{\bff{\omega}}({\bff r}) & = &g_{\omega} [{\bff j}^n({\bff r}) + {\bff j}^p({\bff r})]
\label{KGomegav} \\
\left\{ -\Delta
+m_{\rho}^{2}\right\}
\rho_{0}({\bff r}) & = & g_{\rho} [\rho_v^n({\bff r}) - \rho_v^p({\bff r})],  
 \label{KGRho0} \\
\left\{ -\Delta
+m_{\rho}^{2}\right\}
~{\bff{\rho}}({\bff r}) & = &g_{\rho} [{\bff j}^n({\bff r}) - {\bff j}^p({\bff r})],
\label{KGRhov} \\
-\Delta A_0({\bff r})=e\rho^p_v({\bff r}) , & &  -\Delta {\bff A}({\bff r})=e{\bff j}^p({\bff r}),
\end{eqnarray}
with source terms involving the various nucleonic densities and currents
\begin{eqnarray}
\rho_s^{n,p}({\bff r}) &=& \sum_{i=1}^{N,Z} (\psi_i({\bff r}))^{\dagger}
\hat{\beta} \psi_i({\bff r}),  \\
\rho_v^{n,p}({\bff r}) &=& \sum_{i=1}^{N,Z} (\psi_i({\bff r}))^{\dagger}
\psi_i({\bff r}),  \\
{\bff j}^{n,p}({\bff r}) &=& \sum_{i=1}^{N,Z} (\psi_i({\bff r}))^{\dagger}
\hat{{\bff \alpha}} \psi_i({\bff r}) \label{current-eq}
\end{eqnarray}
where the labels $n$ and $p$ are used for neutrons and protons, respectively. 
In the equations above, the sums run over the occupied positive-energy shell 
model states only ({\it no-sea approximation}) \cite{SW.86,NL1}. Note that the 
spatial components of the vector potential ${\bff A}({\bff r})$ are neglected 
in the calculations since the coupling constant of the electromagnetic 
interaction is small compared with the coupling constants of the meson fields.

  Two terms in the Dirac equation, namely, the Coriolis operator $\hat{J}_x$ 
and the magnetic potential $\bff{V}({\bff r})$ (as well as the currents 
${\bff j}^{n,p}({\bff r})$ in the Klein-Gordon equations) break time-reversal 
symmetry \cite{AR.00}. Their presence leads to the appearance of time-odd mean 
fields. However, one should distinguish time-odd mean fields originating
from Coriolis operator and magnetic potential. The Coriolis operator is always
present in the description of rotating nuclei in the framework of the cranking
model. However, the CRMF calculations, with only these time-odd fields accounted 
for, underestimate the experimental moments of inertia \cite{KR.93,AKR.96}.
A similar situation also holds in nonrelativistic theories \cite{DD.95,YM.00}.
The inclusion of the currents ${\bff j}^{n,p}({\bff r})$ into the Klein-Gordon
equations, which leads to the space-like components of the vector $\omega$ and 
$\rho$ mesons and thus to magnetic potential ${\bff V}({\bff r})$, considerably 
improves the description of experimental moments of inertia. The effect coming 
from the space-like components of the vector mesons is commonly referred to as 
{\it nuclear magnetism} \cite{KR.89} since the magnetic potential has the structure 
of a magnetic field in the Dirac equation.

  Note that time-odd mean fields related to NM are defined through the Lorentz 
invariance \cite{VALR.05} and thus they do not require additional coupling constants: 
the coupling constants of time-even mean fields are used also for time-odd mean 
fields.

  The goal of the current manuscript is to understand the impact of nuclear 
magnetism (NM) on the properties of rotating nuclei. We will use the terms 
{\it nuclear magnetism} and {\it time-odd mean fields} interchangeably 
throughout the manuscript. However, one should keep in mind that the latter 
term is related only to the time-odd mean fields produced by the magnetic 
potential.
  
  Single-particle orbitals are labeled at rotational frequency $\Omega_x=0.0$ MeV 
by $[Nn_z\Lambda]\Omega^{sign}$. $[Nn_z\Lambda]\Omega$ are the asymptotic quantum 
numbers (Nilsson quantum numbers) of the dominant component of the wave function. 
The superscripts {\it sign} to the orbital labels are used sometimes to indicate 
the sign of the signature $r$ for that orbital $(r=\pm i)$. Note that the labelling 
by means of Nilsson labels is performed only when the calculated shape of nuclear 
configuration is prolate or near-prolate.

   Many-particle configurations (further nuclear configurations or configurations) 
are specified by the occupation of available single-particle orbitals. In the 
calculations without pairing, the occupation numbers $n$ are integer ($n=0$ or 1). In 
addition, in the CRMF code it is possible to specify the occupation of either $r=+i$ 
or $r=-i$ signature of the single-particle state. In odd-odd nuclei, all single-particle 
states of specific 
(proton and neutron) subsystem with exception of one are pairwise occupied. We will call 
this occupied single-particle state of fixed signature for which its time-reversal 
(signature) counterpart state is empty as {\it blocked state} in order to simplify the 
discussion.  The specification of nuclear configuration by means of listing all 
occupied single-particle states is unpractical. Thus, in odd-odd nuclei the Nilsson 
labels of the blocked proton and neutron states and their signatures are used for 
configuration labelling.

  The CRMF equations are solved in the basis of an anisotropic three-dimensional 
harmonic oscillator in Cartesian coordinates characterized by the deformation parameters 
$\beta_0$ and $\gamma$ as well as the oscillator frequency $\hbar \omega_0= 41 A^{-1/3}$ 
MeV. Our selection of the deformation parameters of the basis is the same as in earlier
systematic studies of rotating nuclei in different regions of nuclear chart 
(Refs.\ \cite{AKR.96,ALR.98,ARR.99,CRHB,AA.08}). $\gamma=0^{\circ}$ is used in all 
calculations.  The deformation parameter 
$\beta_0$ of the basis is selected in such a way that it provides the convergence to 
the local minimum under study. Thus, $\beta_0=0.25$ is used in the case of normal-deformed 
(ND) states,  $\beta_0=0.2$ and $\beta_0=0.5$ in the case of superdeformed (SD) states in 
the $A\sim 60$ and $A\sim 150,190$ mass regions, respectively, and $\beta_0=1.0$ in 
the case of the hyperdeformed (HD) states. The truncation of basis is performed 
in such a way that all states belonging to the shells up to fermionic  $N_F$=12 and bosonic 
$N_B$=20 are taken into account in the calculations of ND and SD states. The fermionic basis is 
increased up to $N_F$=14 in the calculations of HD states. Numerical analysis indicates that 
these truncation schemes provide sufficient numerical accuracy for the physical quantities of 
interest (see Refs.\ \cite{AKR.96,ALR.98,ARR.99,CRHB,AA.08}).
 
   The majority of the calculations are performed with the NL1 parametrization \cite{NL1} 
of the RMF Lagrangian since this parametrization has been used extensively in earlier 
systematic studies of rotating nuclei across the nuclear chart 
\cite{AKR.96,ALR.98,ARR.99,CRHB,A.08,AA.08}.

  To investigate the impact of NM (time-odd mean fields) on physical observables, 
the CRMF calculations are performed in two calculational schemes for fixed 
configurations:
\begin{itemize}
\item
 Fully self-consistent calculations with NM included (hereafter denoted NM calculations),
which take into account space-like components of the vector mesons (Eqs.\ (\ref{KGomegav}), 
(\ref{KGRhov}) and (\ref{magnetic})), currents (Eqs.\ (\ref{KGomegav}), (\ref{KGRhov}), 
and (\ref{current-eq})), and magnetic potential {\bff V}({\bff r}) (Eq.\ (\ref{magnetic})).

\item
  Fully self-consistent calculations without NM (hereafter denoted as WNM calculations), 
which omit space-like components of the vector mesons (Eqs.\ (\ref{KGomegav}), 
(\ref{KGRhov}) and (\ref{magnetic})), currents in the Klein-Gordon equations (Eqs.\ 
(\ref{KGomegav}) and (\ref{KGRhov}), and magnetic potential {\bff V}({\bff r}) 
(Eq.\ (\ref{magnetic})). The results of the NM and WNM calculations are always compared 
for the same nuclear configuration.

\end{itemize}

 These are the ways in which the effects of time-odd mean fields can be studied, and as 
such they are frequently used in DFT studies of rotating systems, both in relativistic 
and non-relativistic frameworks \cite{KR.93,DD.95,YM.00,AR.00,A.08}. One should, however, 
keep in mind that if time-odd fields are neglected, the local Lorentz  invariance (Galilean 
invariance in non-relativistic framework \cite{DD.95,CDK.08}) is violated. The inclusion 
of time-odd mean fields restores the Lorentz invariance.
   
  It is interesting to compare the basic features such as Lorentz invariance and the 
definition of the coupling constants of the time-odd channel of the CDF theory discussed
above with the ones of non-relativistic Skyrme energy density functional (EDF) theory. 
It was recognized in earlier Skyrme DFT studies, that the connection between the coupling 
constants of time-odd and time-even channels depends on what entity, namely, Skyrme 
force or energy density functional is considered 
to be more fundamental \cite{DD.95,CDK.08,DCK.10}. If the Skyrme force is considered 
more fundamental then the time-odd constants are determined as a function of time-even 
constants \cite{DD.95,DCK.10}.  However, since the time-even coupling constants 
are usually adjusted solely to the time-even observables, the resulting values of the 
time-odd coupling constants simply ``fictitious''  or ``illusory'', as noted already 
in Ref.\ \cite{N.75} (see also Ref.\ \cite{DCK.10}). On the contrary, in the framework
of the Skyrme energy density functional theory, time-odd properties of the functional 
are independent of time-even properties which is a consequence of broken link between 
the Skyrme force and the density functional.

   The question of whether Galliean invariance must be imposed in Skyrme EDF is not
yet resolved \cite{CDK.08}, despite the fact that it is imposed in many studies. Note
that in many phenomenological approaches, such as the noninteracting or interacting 
shell models, Galilean symmetry is not considered, because the translational motion is 
not within the scope of such models \cite{CDK.08}.

 It is also important to mention that the cranking models based on phenomelogical 
Woods-Saxon or Nilsson potentials do not incorporate time-odd mean fields. However, 
they succesfully describe rotating nuclei \cite{AFLR.99,SW.05}.
  
  Note that the Coriolis term is present in NM and WNM calculations. This means 
that the currents (Eq.\ (\ref{current-eq})) are always present in rotating nuclei. 
However, it is important to distinguish the currents induced by the Coriolis term and 
the ones which appear due to magnetic potential. The currents, which appear in the
WNM calculations, are generated by the Coriolis term. Thus, we will call them as 
{\it Coriolis induced currents}.  On the contrary, the currents in the NM calculations
are generated by both the Coriolis term and magnetic potential. The difference of the
currents in the NM and WNM calculations is attributable to magnetic potential.
Thus,  the currents $[{\bff j}^{n,p}({\bff r})]^{NM} - [{\bff j}^{n,p}({\bff r})]^{WNM}$
will be called {\it magnetic potential induced currents}.
 
   In the following, the contribution $\Delta O^{NM-contr}$ (in percentage) of NM 
to the physical observable $O$ is defined as
\begin{eqnarray}
\Delta O^{NM-contr} = \frac{O^{NM}-O^{WNM}}{O^{NM}}\times 100\%.
\end{eqnarray} 
The physical observables, most frequently used in the analysis of rotating
nuclei, are kinematic ($J^{(1)}$) and dynamic ($J^{(2)}$) moments of inertia 
which are defined as
\begin{eqnarray}
J^{(1)}(\Omega_x)=\frac{J}{\Omega_x},\qquad 
J^{(2)}(\Omega_x)=\frac{dJ}{d\Omega_x}
\label{Jmoments}
\end{eqnarray}
where $J$ is the expectation value of the total angular momentum along 
the $x$-axis. In the CRMF theory, this quantity is defined 
as a sum of the expectation values of the single-particle angular momentum 
operators $\hat{\jmath}_{x}$ of the occupied states
\begin{eqnarray}
J=\sum_{i}\langle i|\hat{\jmath}_{x}|i\rangle .
\label{Jmicro}
\end{eqnarray}
Thus, the modifications of the moments of inertia due to NM can be traced back to the 
changes of the single-particle expectation values 
$\langle \hat{\jmath}_{x}\rangle _{i}=$ $\langle i|\hat{\jmath}_{x}|i\rangle $ 
and the corresponding contributions of spin ($\langle \hat{s}_{x}\rangle _{i}$) 
and orbital ($\langle \hat{l}_{x}\rangle _{i}$) angular momenta \cite{AR.00}.

%%%%%%%%%%%%%%%%%%%%%%%%%%%%%%%%%%%%%%%%%%%%%%%%%%%%%%%%%%%%%%%%%%%%%%%%%%%
\section{Nuclear magnetism and band crossing features}
\label{Sect-band-crossing}
%%%%%%%%%%%%%%%%%%%%%%%%%%%%%%%%%%%%%%%%%%%%%%%%%%%%%%%%%%%%%%%%%%%%%%%%%%%

%%%%%%%%%%%%%%%%%%%%%%%%%%%%%%%%%%%%%%%%%%%%%%%%%%%%%%%%%%%%%%%%%%%%%%%%%%%%%%%%%
\begin{figure}
\centering
\includegraphics[width=8.0cm]{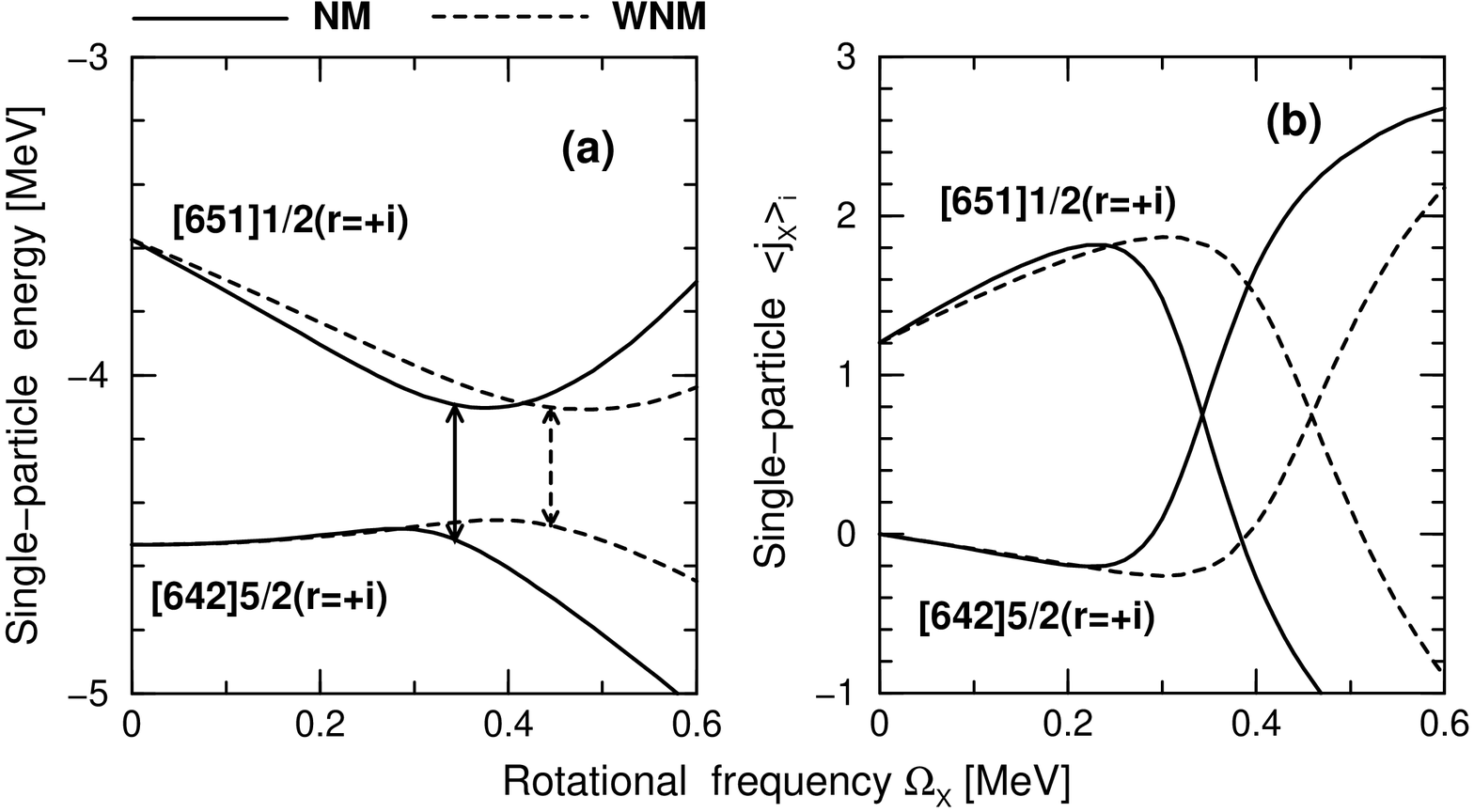}
\caption{(a) Proton single-particle energies (Routhians)
in the self-consistent rotating potential as a function of
rotational frequency $\Omega_x$ obtained in the CRMF
calculations with and without NM. They are given along the
deformation path of the lowest SD configuration in $^{194}$Pb.
Only interacting $[651]1/2^+$ and $[642]5/2^+$ orbitals
are shown, see Fig.\ 1 in Ref.\ \protect\cite{CRHB} for full
spectra. (b) The expectation values $\langle \hat{j}_x \rangle_i$
of the single-particle angular momentum operator $\hat{j}_x$ of the
orbitals shown on panel (a). Solid and dashed arrows are used to 
indicate the frequencies (as well as the energy gap between the 
interacting orbitals in panel (a)) at which the band crossings take 
place  in the calculations with and without NM, respectively.
\label{espe-jx-np}}
\end{figure}
%%%%%%%%%%%%%%%%%%%%%%%%%%%%%%%%%%%%%%%%%%%%%%%%%%%%%%%%%%%%%%%%%%%%%%%%%%%%%%%%

%%%%%%%%%%%%%%%%%%%%%%%%%%%%%%%%%%%%%%%%%%%%%%%%%%%%%%%%%%%%%%%%%%%%%%%%%%%%%%%%%
\begin{figure}
\centering
\includegraphics[width=8.0cm]{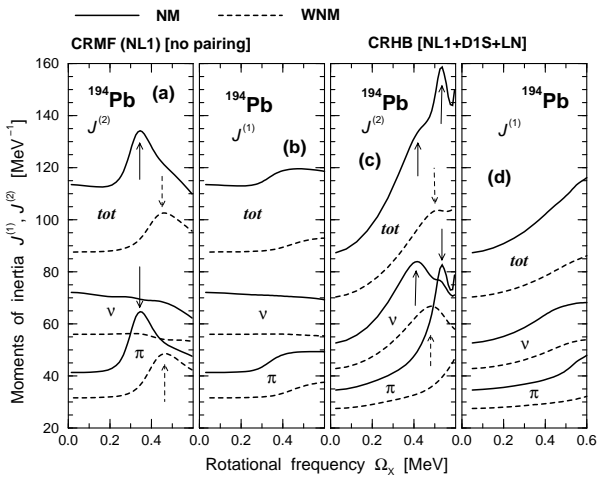}
\caption{Kinematic ($J^{(1)}$) and dynamic ($J^{(2)}$) moments of inertia for the 
lowest SD configuration in $^{194}$Pb obtained in the calculations with and without 
NM. Proton and neutron contributions to these quantities are indicated by $\pi$ and 
$\nu$, while total moments by '$tot$'. Panels (a) and (b) show the results obtained in 
the calculations without pairing, while panels (c) and (d) show the results of the 
calculations within the CRHB+LN framework. Solid and dashed arrows are used to indicate 
the frequencies at which the band crossings take place in the calculations with and 
without NM, respectively.
\label{pb194-crossing}}
\end{figure}
%%%%%%%%%%%%%%%%%%%%%%%%%%%%%%%%%%%%%%%%%%%%%%%%%%%%%%%%%%%%%%%%%%%%%%%%%%%%%%%%

  Since NM substantially modifies  the single-particle properties (energies, alignments) 
\cite{AR.00,VALR.05}, it is reasonable to expect that the band crossing features are 
affected by NM. In order to study this question, the CRMF (without pairing) and the 
CRHB+LN calculations have been performed for lowest superdeformed (SD) band in 
$^{194}$Pb. In the CRHB+LN calculations, the D1S Gogny force \cite{D1S} is used in 
pairing channel and  an approximate particle number projection is performed by 
means of the Lipkin-Nogami method \cite{CRHB}.

  The unpaired proton band crossing seen in the CRMF calculations  originates from 
the interaction between the $\pi[642]5/2^+$ 
and $\pi[651]1/2^+$ orbitals (Fig.\ \ref{espe-jx-np}a). Since NM increases 
somewhat the single-particle alignment $\langle \hat{j}_x \rangle_i$ (Fig.\ 
\ref{espe-jx-np}b) and the slope of the routhian for the $\pi[651]1/2^+$ orbital 
(Fig.\ \ref{espe-jx-np}a), the band crossing takes place at lower frequency. The shift 
of crossing frequency due to NM is considerable (120 keV) from 0.465 MeV (WNM) down 
to 0.345 MeV (NM), Fig.\ \ref{espe-jx-np}a. The calculations also suggest that 
the strength of the interaction between two interacting orbitals at the band 
crossing is modified in the presence of NM as seen in the change of the energy 
distance (gap) between these two orbitals at the crossing frequency (Fig.\ 
\ref{espe-jx-np}a).

   An additional mechanism affecting the band crossing frequencies will be active
in odd- and odd-odd mass nuclei as well as in excited configurations of even-even 
nuclei. In such configurations, there is at least one single-particle state the 
opposite signature of which is not occupied. This results in the currents at 
$\Omega_x=0.0$ MeV \cite{AA.10}. The energy  splitting between different signatures 
of the single-particle states at no rotation is a typical consequence of these 
currents (see Sec.\ IVA in Ref.\ \cite{AA.10} for  more details).  As a result, 
the energy gap between interacting orbitals  at $\Omega_x=0.0$ MeV can become larger 
or smaller dependent on the impact of the currents on the single-particle energies 
of interacting states. Consequently, this change in the energy gap will translate 
into higher or lower band crossing frequencies. Note that for simplicity we assume 
that the $\Omega_x=0.0$ currents will not modify the alignment properties of 
interacting orbitals; this translates into the independence of the single-particle
routhian slope in the energy versus $\Omega_x$ plot (see, for example, Fig.\ 
\ref{espe-jx-np}a) on the $\Omega_x=0.0$ currents.

  Fig.\ \ref{espe-jx-np}a can be used to illustrate this mechanism. 
Let assume that the $\Omega_x=0.0$ currents will increase the energy gap 
between the $\pi[642]5/2^+$ and $\pi[651]1/2^+$ orbitals at $\Omega_x=0.0$ MeV: 
this will lead to higher band crossing frequencies. However, the band crossing 
frequencies will decrease in the case when the energy gap between these orbitals 
at $\Omega_x=0.0$ MeV becomes smaller in the presence of the $\Omega_x=0.0$ currents. 
The assumption that the $\Omega_x=0.0$ currents do not have an impact on the 
alignment properties of interacting orbitals is definitely too simplistic but it 
allows to illustrate the fact that NM can both decrease and increase the band 
crossing frequencies.

  This mechanism is not active in the configuration of even-even $^{194}$Pb 
nucleus discussed above since both signatures of all states below the Fermi level 
are pairwise occupied.  As a result, no current is present at $\Omega_x=0.0$ MeV.

  The impact of NM on band crossing features is also seen in the CRHB+LN calculations 
where the alignment of the pairs of $j_{15/2}$ neutrons and $i_{13/2}$ protons causes 
the shoulder and peak in total dynamic moment of inertia $J^{(2)}$ (Fig.\ 
\ref{pb194-crossing}c) (see also Ref.\ \cite{CRHB}). Note that each of these two 
alignments creates a peak in the dynamic moment of inertia of corresponding subsystem. 
NM shifts the paired neutron band crossing to lower frequencies by 70 keV from $0.485$ 
MeV (WNM) to $0.415$ MeV (NM). Paired proton band crossing lies in the calculations 
with NM at $\Omega_x=0.535$ MeV, while only the beginning of this crossing is seen 
in the calculations without NM (Fig.\ \ref{pb194-crossing}c).

\begin{figure}
\centering
\includegraphics[width=8.0cm]{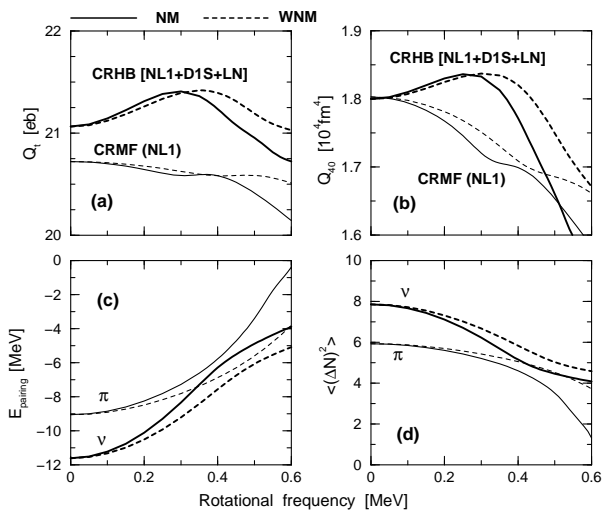}
\caption{Transition quadrupole moments $Q_t$ (panel (a)), mass hexadecapole 
moments $Q_{40}$ (panel (b)), proton and neutron pairing energies 
$E_{pairing}=-1/2\,\,Tr(\Delta \kappa)$
(panel (c)) and proton and neutron particle number
fluctuations $\langle (\Delta \hat{N})^2 \rangle$
(panel (d)) of the lowest SD solution in $^{194}$Pb obtained
in the calculations with and without NM. Thick and thin lines
in the upper panels are used for the results obtained in the
CRHB and CRMF theories, respectively. On the bottom panels,
thick and thin lines are used for neutron and proton quantities,
respectively.
\label{pb94-obs-comp}}
\end{figure}

  The origin of this effect is twofold. Similar to the  unpaired calculations, 
the part of it can be traced to the fact that NM increases the expectation values
$\langle \hat{j}_x \rangle_i$ of the orbitals located at the bottom of the shell 
(the discussed orbitals are of this kind) \cite{AR.00}. The corresponding larger 
slope of the quasiparticle routhians causes the shift of the crossing to lower 
frequencies. However, an additional contribution comes from the modification of the 
pairing  by NM. There is a difference in the pairing energies calculated with and 
without NM which increases with rotational frequency, see Fig.\ \ref{pb94-obs-comp}c,d. 
The pairing in the calculations with NM is weaker.   This can be  explained by
the increase of $\langle \hat{j}_x \rangle_i$ of the orbitals located at the bottom 
of the shell due to NM (see above). The gradual breaking of high-$j$ pairs proceeds
faster, which is reflected in a faster decrease of pairing with increasing $\Omega_x$.
Thus we can specify this effect as {\it an anti-pairing effect induced by NM}. 

%%%%%%%%%%%%%%%%%%%%%%%%%%%%%%%%%%%%%%%%%%%%%%%%%%%%%%%%%%%%%%%%%%%%%%%%%%%%%%%%%
%%%%%%%%%%%%%%%%%%%%%%%%%%%%%%%%%%%%%%%%%%%%%%%%%%%%%%%%%%%%%%%%%%%%%%%%%%%%%%%%

  These considerable differences in the crossing frequencies obtained in the 
calculations with and without NM cannot be attributed to the differences in 
equilibrium deformations, since calculated transition quadrupole moments 
$Q_t$ and mass hexadecapole moments $Q_{40}$ obtained in the calculations with
and without NM differ only marginally before band crossing, see Figs.\ 
\ref{pb94-obs-comp}a,b.

  The influence of time-odd mean fields on band crossing features has been studied 
by means of a schematic non-self-consistent model based on the Skyrme forces in 
Ref.\ \cite{PWM.85}. In this study, time-odd fields emerging from the 
${\bf S}^2$ and $-{\bf S} \Delta {\bf S}$ terms of the Skyrme Hamiltonian shift 
the alignment of the $i_{13/2}$ neutron pair to higher frequencies in $^{158}$Dy. 
On the contrary, this crossing appears at lower frequencies in the CRHB+LN calculations 
when NM is taken into account. This difference is not surprising considering the fact 
that time-odd mean fields are not well defined in non-relativistic density functional 
theories \cite{DD.95,SDMMNSS.10}. It was also suggested in the cranked Skyrme 
Hartree-Fock framework that time-odd mean fields may be responsible for band 
crossing in yrast superdeformed band of $^{60}$Zn \cite{DDW.03}. However, this 
crossing is described as paired band crossing  in the CRHB+LN  calculations 
\cite{Pingst-A30-60}.
 
  Above discussed CRMF and CRHB+LN examples clearly show that the modifications of 
band crossing  features (crossing frequencies and the features of the kinematic 
and dynamic moments of inertia in band crossing region) caused by NM are substantial 
and depend on the underlying modifications of single-particle properties such 
as alignments and single-particle (quasi-particle) energies.

%%%%%%%%%%%%%%%%%%%%%%%%%%%%%%%%%%%%%%%%%%%%%%%%%%%%%%%%%%%%%%%%%%%%%
\section{Particle number and deformation  dependences of the impact 
of nuclear magnetism on the moments of inertia}
\label{pndefdep-mom}
%%%%%%%%%%%%%%%%%%%%%%%%%%%%%%%%%%%%%%%%%%%%%%%%%%%%%%%%%%%%%%%%%%%%%

   In the current section, the particle number and deformation dependence 
of the impact of NM on the kinematic moments of inertia are discussed in 
detail. We consider the contribution of NM to kinematic moment of inertia, 
namely, the $(J^{(1)}_{NM}-J^{(1)}_{WNM})/J^{(1)}_{NM}$ quantity, and its 
variations as a function of particle number and deformation.  In addition, 
we investigate how close fully self-consistent value of the kinematic moment 
of inertia comes to the rigid body moment of inertia $J_{rig}$. The latter 
quantity is obtained in one-dimensional cranking approximation with the 
rotation defined around the $x$-axis from the calculated density distribution 
$\rho({\bf r})$ by 
\begin{eqnarray}
J_{rig}=\int \rho({\bf r}) (y^2+z^2)d^3r
\label{Jrigid}
\end{eqnarray}
%
%%%%%%%%%%%%%%%%%%%%%%%%%%%%%%%%%%%%%%%%%%%%%%%%%%%%%%%%%%%%%%%%%%%%%%%%%%%%%%%%%
\begin{figure}
\centering
\includegraphics[width=8.0cm]{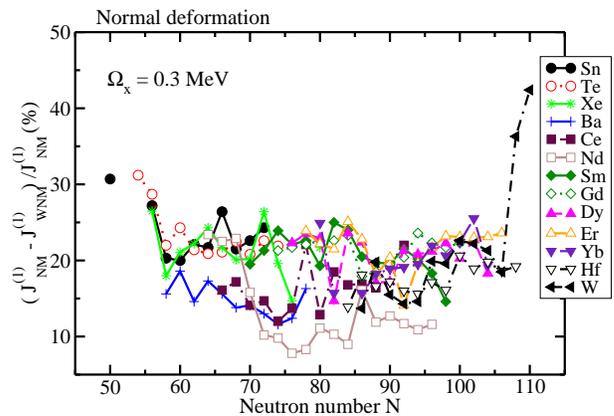}
\vspace{0.8cm}
\caption{ (Color online)  The contribution (in \%) of NM to the kinematic moments 
of inertia of nuclei in different isotope chains at normal deformation. The results 
for the lowest in energy solutions are shown at rotational frequency $\Omega_x=0.3$ 
MeV. The frequency has been fixed at this value in order to make the comparison with 
the results of Ref.\ \cite{DFPCU.04} straightforward.
\label{J1-sys-ND}}
\end{figure}
%%%%%%%%%%%%%%%%%%%%%%%%%%%%%%%%%%%%%%%%%%%%%%%%%%%%%%%%%%%%%%%%%%%%%%%%%%%%%%%%
%
 The contributions of NM to kinematic moment of inertia (the $(J^{(1)}_{NM}-J^{(1)}_{WNM})/J^{(1)}_{NM}$
quantity) for normal deformed bands in a number of isotope chains with proton number $Z\geq 50$ 
are shown as a function of neutron number in Fig.\ \ref{J1-sys-ND}. Only the cases in which
the nuclear configurations are the same in the calculations with and without nuclear magnetism
are shown in this figure. NM typically increases the calculated kinematic moments of inertia by 
10-30\%. However, this increase is around 40\% in the $N=108,110$ W isotopes. Considerable 
fluctuations of the $(J^{(1)}_{NM}-J^{(1)}_{WNM})/J^{(1)}_{NM}$ quantity as a function of neutron 
number seen in some isotope chains are due to the changes in underlying single-particle structure. Large 
changes in the $(J^{(1)}_{NM}-J^{(1)}_{WNM})/J^{(1)}_{NM}$ quantity are seen on going 
from the isotope with neutron number $N$ to the isotope with $N+2$ when two neutron 
single-particle orbitals, by which the configurations of compared nuclei differ,
have the expectation values  of the single-particle angular momenta $\langle \hat{j}_x \rangle_i$ 
strongly affected by NM. The opposite is also true when two neutron single-particle 
orbitals, by which the configurations of compared nuclei differ, have the expectation values  
of the single-particle angular momenta $\langle \hat{j}_x \rangle_i$ that are only 
marginally affected by NM. Note that in some cases proton configurations of two neighboring 
nuclei with neutron numbers $N$ and $N+2$ are also different due to the deformation changes; 
this also contributes into the fluctuations of the $(J^{(1)}_{NM}-J^{(1)}_{WNM})/J^{(1)}_{NM}$ 
quantity as a function of neutron number.

  One can also extract from Fig.\ \ref{J1-sys-ND} the dependence of the contributions of 
NM to kinematic moments of inertia on proton number $Z$ by considering the results of 
the calculations at constant value of neutron number $N$. Such analysis reveals the 
fluctuations in the $(J^{(1)}_{NM}-J^{(1)}_{WNM})/J^{(1)}_{NM}$ quantities which are similar 
to the ones discussed above. The origin of these fluctuations can again be traced back 
to the changes (as a function of proton number) in underlying  single-particle 
structure.

\begin{figure}
\centering
\vspace{1.2cm}
\includegraphics[width=8.0cm]{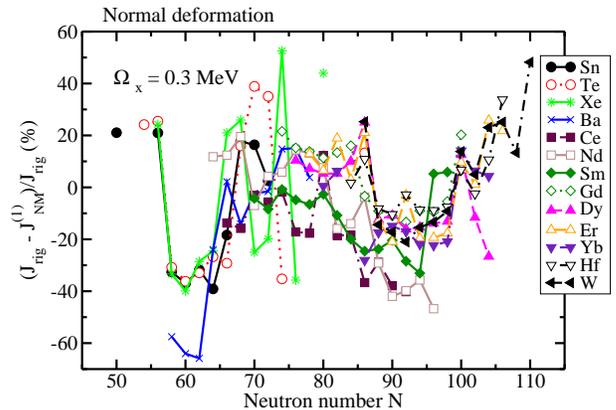}
\caption{ (Color online) The difference (in \%) between the rigid body moments 
of inertia $J_{rig}$ and kinematic moments of  inertia calculated with NM 
for the nuclear configurations shown in Fig.\ \ref{J1-sys-ND}.
\label{J1-sys-ND-class}}
\end{figure}

  Fig.\ \ref{J1-sys-ND-class} compares rigid body moments of inertia $J_{rig}$ (Eq.\ (\ref{Jrigid}))
with fully microscopic kinematic moments of inertia $J^{(1)}_{NM}$ (Eqs.\ (\ref{Jmoments}) and 
(\ref{Jmicro})) obtained in the calculations with NM using the $(J_{rig}-J^{(1)}_{NM})/J_{rig}$ 
quantity. One can see that considerable deviations (in majority of the cases being in the window 
of $\pm 30\%$ but reaching $\pm 60\%$ in some nuclei) between these two moments of inertia are 
observed at normal deformation.

%%%%%%%%%%%%%%%%%%%%%%%%%%%%%%%%%%%%%%%%%%%%%%%%%%%%%%%%%%%%%%%%%%%%%%%%%%%%%%%%%
%%%%%%%%%%%%%%%%%%%%%%%%%%%%%%%%%%%%%%%%%%%%%%%%%%%%%%%%%%%%%%%%%%%%%%%%%%%%%%%%

  The analysis within the framework of the periodic orbit theory \cite{DFPCU.04} concluded 
that the deviations of the moments of inertia from the rigid-body value at high spin are 
determined by the shell structure of a system of independent fermions confined by a leptodermous 
potential. For the case of prolate deformation and the rotation perpendicular to the symmetry 
axis (the majority of the cases studied in the current manuscript fall under this category), 
the meridian orbits determine the shell moments of inertia because only they enclose 
rotational flux \cite{DFPCU.04}.

  Large similarities are seen between the results of our calculations and the ones based 
on the cranked Woods-Saxon potential in Ref.\ \cite{DFPCU.04}. For example, right bottom 
panel in Fig.\ 10 of Ref.\ \cite{DFPCU.04} shows the difference $J_{pper}-J_{rig}$ between 
the moments of inertia $J_{pper}$ calculated in the cranked Woods-Saxon potential and 
rigid-body moments of inertia $J_{rig}$ for the case of prolate deformation and the rotation 
around the axis perpendicular to the symmetry axis. If one corrects for the difference in 
the representation of calculated quantities [($J_{pper}-J_{rig}$) in Ref.\ \cite{DFPCU.04} 
and $(J_{rig}-J^{(1)}_{NM})/J_{rig}$ in the present manuscript], then one can see that our 
results show similar shell dependence of the $(J^{(1)}_{NM}-J^{1}_{rig})$ quantities as 
the one seen in Fig.\  10 of Ref.\  \cite{DFPCU.04}. Some differences between these two 
calculations are in part due to simplistic method of the calculation of the rigid-body 
moments of inertia in Ref.\ \cite{DFPCU.04} (see Sec. IIB of Ref.\ \cite{DFPCU.04} for details).

%%%%%%%%%%%%%%%%%%%%%%%%%%%%%%%%%%%%%%%%%%%%%%%%%%%%%%%%%%%%%%%%%%%%%%%%%%%
\begin{figure}
\centering
\includegraphics[width=8.0cm]{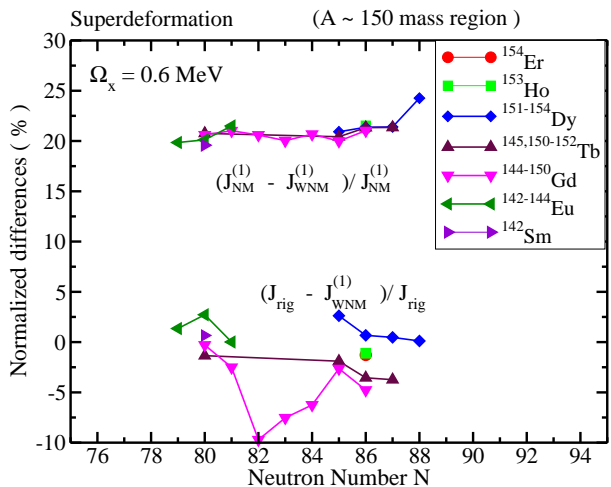}
\caption{ (Color online) The contribution $(J_{NM}^{(1)}-J_{WNM}^{(1)})/J_{NM}^{(1)})$ 
(in \%) of NM to the kinematic moments of inertia and the difference 
$(J_{rig}-J_{NM}^{(1)})/J_{rig}$ (in \%) between the rigid body moments of inertia and 
kinematic moments of inertia calculated with inclusion of NM. The results for yrast 
SD configurations in the $A\sim 150$ mass region of superdeformation are 
shown at rotational frequency $\Omega_x=0.6$ MeV for nuclei which were previously 
analyzed in the CRMF calculations in Refs.\ \cite{AKR.96,ALR.98}.
\label{J1-sys-SD}}
\end{figure}
%%%%%%%%%%%%%%%%%%%%%%%%%%%%%%%%%%%%%%%%%%%%%%%%%%%%%%%%%%%%%%%%%%%%%%%%%%%%%%%%%

  The CRMF calculations describe rather well the kinematic moments of inertia of 
normal-deformed  \cite{AF.05,NL3*}  and smooth-terminating \cite{VALR.05,NL3*} bands 
at high spin where the pairing is negligible. Experimental data on kinematic moments 
of inertia of normal-deformed rotational bands at low spin [which are strongly affected 
by pairing] are also well described in the cranked relativistic Hartree-Bogoliubov 
calculations \cite{AKRRE.00,A250}. These results together with the ones presented in 
the current manuscript strongly support the conclusion that weakly- and normal-deformed 
nuclei show the moments of inertia which strongly deviate from the rigid-body value 
(see also Refs.\ \cite{AFLR.99,DFPCU.04}).

\begin{figure}
\centering
\includegraphics[width=8.0cm]{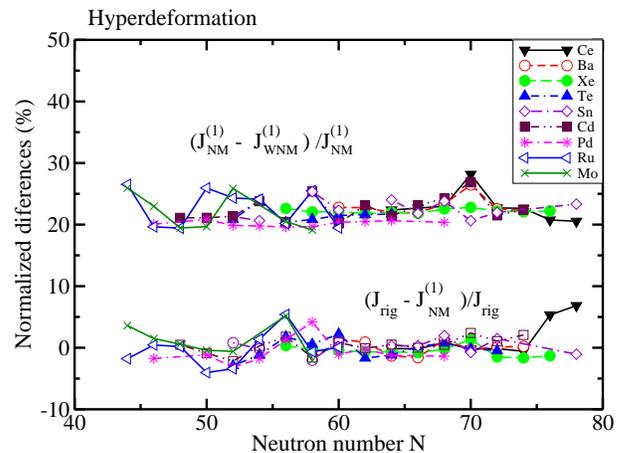}
\caption{(Color online) The same as in Fig.\ \ref{J1-sys-SD} but for the hyperdeformed 
configurations. The results of the calculations are shown for the yrast HD configurations, 
studied in the systematic survey of the hyperdeformation in the $Z=40-58$ region of the 
nuclear chart \cite{AA.08,AA.09}, at spins at which they become yrast (rotational 
frequency $\Omega_x\approx 0.8-1.0$ MeV).
\label{J1-sys-HD}}
\end{figure}

  Figs.\ \ref{J1-sys-SD} and \ref{J1-sys-HD} show the results of calculations 
for yrast SD configurations in the $A\sim 150$ mass region of superdeformation 
and for yrast hyperdeformed (HD) configurations in the $Z=40-58$ part of the nuclear 
chart, respectively. It is clearly seen that the $(J^{(1)}_{NM}-J^{(1)}_{WNM})/J^{(1)}_{NM}$ 
and $(J_{rig}-J^{(1)}_{NM})/J_{rig}$ quantities at these extreme deformations show much 
smaller fluctuations than the ones at normal deformation. Indeed, the contribution of NM 
into kinematic moment of inertia at SD and HD is in narrow $20-27\%$ range (Figs.\ \ref{J1-sys-SD} 
and \ref{J1-sys-HD}), while it covers much large $9-43\%$ range at normal 
deformation (Fig.\ \ref{J1-sys-ND}). In addition, the values of kinematic moment of inertia calculated 
with NM are typically within 5\% of the rigid body value for the moment of inertia at SD and HD (Figs.\ 
\ref{J1-sys-SD} and \ref{J1-sys-HD}), while much larger fluctuations (typically within 40\% of the rigid 
body value) are seen in the case of normal deformation (Fig.\ \ref{J1-sys-ND-class}).

  Microscopic origin of these features can be traced back to the underlying shell structure. The 
analysis within the periodic orbit theory \cite{DFPCU.04} shows that the single-particle orbits that 
cause shell structure of prolate superdeformed nuclei do not carry rotational flux if the axis of 
rotation is perpendicular to the symmetry axis. Therefore, the moments of inertia of the SD bands in 
such nuclei should be equal to the rigid body value \cite{DFPCU.04}. Such conclusion is in general 
supported by our microscopic calculations which show that the calculated moments of inertia are 
typically within 5\% of rigid-body value. The experimental deviations (obtained under spin assignments 
of Refs.\ \cite{Rag.93,ALR.98}) from the rigid-body values are about 6\% or less in the $A\sim 150$
region of superdeformation (see Fig.\ 5 in Ref.\ \cite{DFPCU.04}).

  We also expect that similar mechanism is also responsible for the observed features of the 
moments of inertia at HD. However, the periodic orbit theory analysis of such features is not 
available and it goes beyond the scope of the current manuscript.

%%%%%%%%%%%%%%%%%%%%%%%%%%%%%%%%%%%%%%%%%%%%%%%%%%%%%%%%%%%%%%%%%%%%%%%%%%%%%%%%%
%%%%%%%%%%%%%%%%%%%%%%%%%%%%%%%%%%%%%%%%%%%%%%%%%%%%%%%%%%%%%%%%%%%%%%%%%%%%%%%%

%%%%%%%%%%%%%%%%%%%%%%%%%%%%%%%%%%%%%%%%%%%%%%%%%%%%%%%%%%%%%%%%%%%%%%%%%%%%%%%
\section{Currents in intrinsic (rotating) frame of collectively rotating nuclei}
\label{Sec-currents}
%%%%%%%%%%%%%%%%%%%%%%%%%%%%%%%%%%%%%%%%%%%%%%%%%%%%%%%%%%%%%%%%%%%%%%%%%%%%%%%

%%%%%%%%%%%%%%%%%%%%%%%%%%%%%%%%%%%%%%%%%%%%%%%%%%%%%%%%%%%%%%%%%%%%%%%%%%%%%%%%%
\begin{figure*}
\centering
\includegraphics[width=16.0cm]{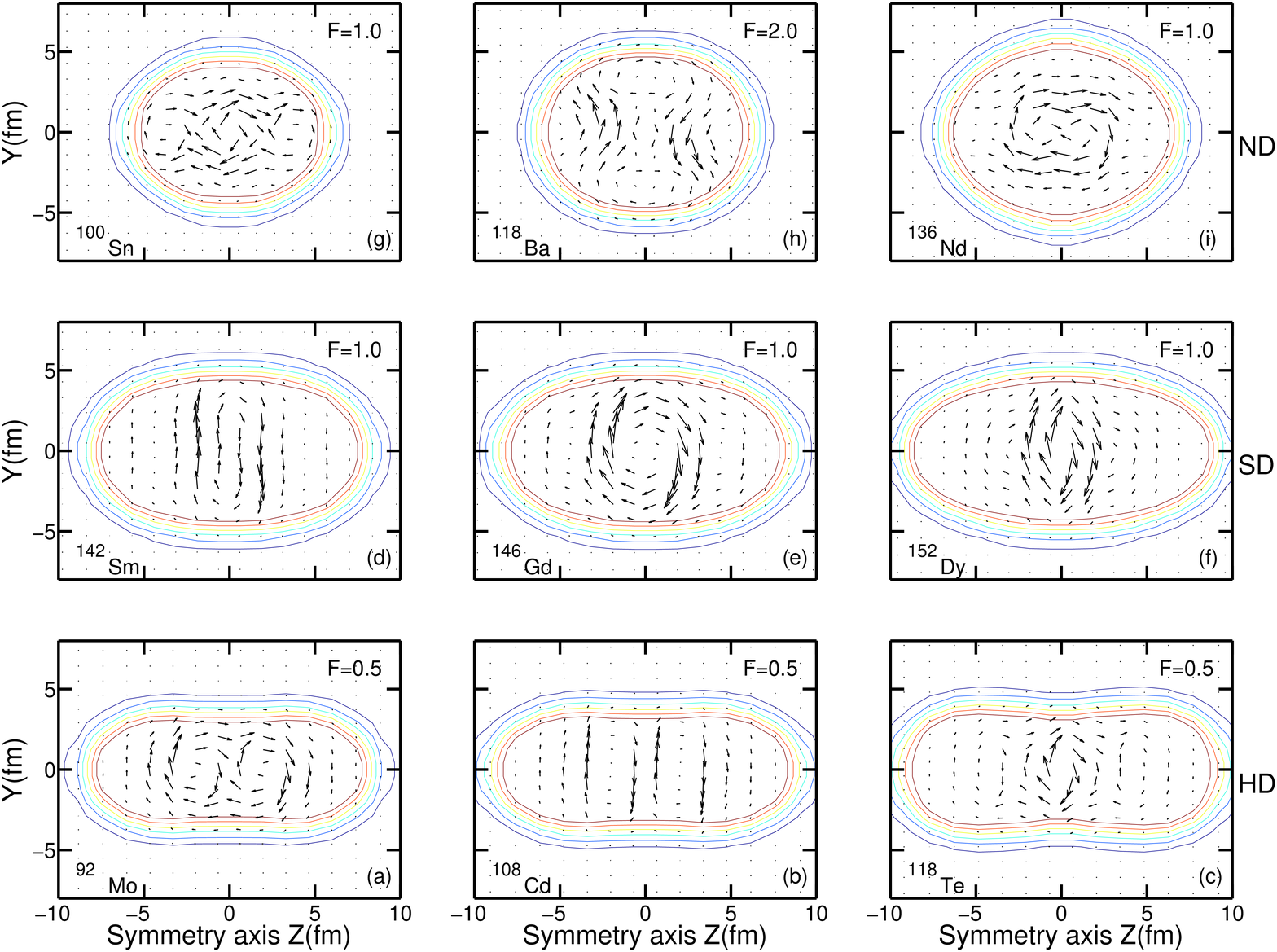}
\caption{ (Color online) Total neutron current distributions  {\bff j}$^n$({\bff r}) 
in the intrinsic frame in the $y-z$ plane for several normal-deformed
(ND) (upper row), superdeformed (SD) (middle row) and hyperdeformed (HD) (bottom row)
configurations in different nuclei. They are shown at $x \approx 0.48$ fm in the
case of ND and SD configurations, and at $x \approx 0.42$ fm in the case of HD
configurations. The results of calculations are shown at rotational frequencies 
$\Omega_x=0.3$ MeV and $\Omega_x=0.5$ MeV for the ND and SD configurations,  respectively, 
and at the spin ($\Omega_x\sim 1.0$ MeV) at which the HD configurations 
become yrast (see Ref.\ \cite{AA.08} for details) in the case of HD configurations. 
The currents in panels (d-g) and (i) are plotted  at arbitrary units for better 
visualization. The currents in other panels are normalized to the currents in panels 
(d-g) and (i) by using factor F. This factor is chosen  in such way that the current 
distribution for every nucleus is clearly seen. The shape and size of the nucleus are 
indicated by density lines which are plotted in the range $0.01-0.06$ fm$^{-3}$ in 
step of 0.01 fm$^{-3}$.
\label{currents}}
\end{figure*}
%%%%%%%%%%%%%%%%%%%%%%%%%%%%%%%%%%%%%%%%%%%%%%%%%%%%%%%%%%%%%%%%%%%%%%%%%%%%%%%%

  Current distributions in the intrinsic (rotating) frame have been studied earlier 
in several publications. It is well known that there are no currents in the
intrinsic frame if the rigid non-spherical body rotates uniformly (rigid rotation) 
(see Sec.\ 6A-5 in Ref.\ \cite{BM.book}).  The general aspects of the velocity (current)
fields have been discussed in detail in the framework of single-particle Schr{\"o}dinger
fluid \cite{KG.77}, which exhibits a remarkably rich variety of fluid dynamical features, 
including compressible flow and line vortices. Nuclear intrinsic vorticity and its coupling 
to global rotations have been studied within the so-called routhian approach both in 
semiclassical approach \cite{DSK.85,MQS.97} and in fully self-consistent cranked 
Hartree-Fock and Hartree-Fock-Bogoliubov approaches based on the Skyrme force \cite{LSQM.03}. 
The current distributions in rotating frame have been studied in phenomenological cranking 
approaches based on harmonic oscillator \cite{R.76,GR.78-0,GR.78,DSK.85} and  Nilsson 
\cite{KM.79} potentials  and in self-consistent cranking approaches based on the Skyrme force 
\cite{FKMW.80,LSQM.03}. Note that the intrinsic current field (as any vector field according 
to the Hemholtz's theorem) can be split into irrotational and intrinsic vortical fields 
\cite{MQS.97}.

%%%%%%%%%%%%%%%%%%%%%%%%%%%%%%%%%%%%%%%%%%%%%%%%%%%%%%%%%%%%%%%%%%%%%%%%%%%%%%%%%
\begin{figure*}
\centering
\includegraphics[width=16.0cm]{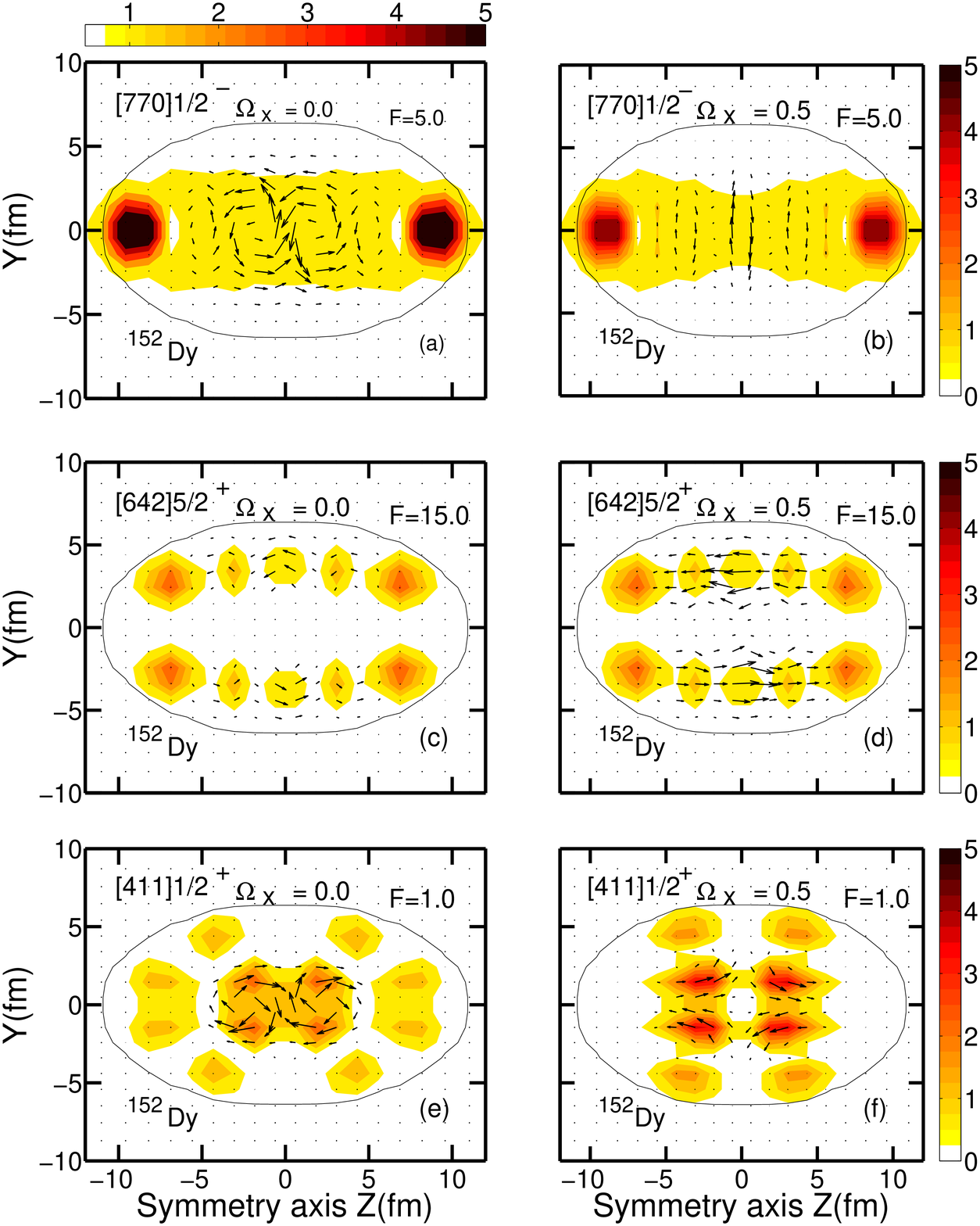}
\caption{(Color online) Current distributions {\bff j}$^n$({\bff r}) produced by 
single neutron in indicated single-particle states of the yrast SD configuration in 
$^{152}$Dy at rotational frequencies $\Omega_x=0.0$ MeV (left panels) and $\Omega_x=0.5$ 
MeV (right panels). The shape and size of the nucleus are indicated by density line 
which is plotted at $\rho=0.01$ fm$^{-3}$. The currents in panels (e),(f) are plotted at 
arbitrary units for better visualization. The currents in other panels are normalized to 
the currents in panels (e) and (f) by using factor F.  The currents and densities are shown 
in the intrinsic frame in the $y-z$ plane at $x=0.48$ fm. The single-neutron density 
distributions due to the occupation of the indicated Nilsson state are shown by colormap. 
Note that slightly different colormap is used in panel (a) for better visualization of 
densities.
\label{sp-currents}}
\end{figure*}
%%%%%%%%%%%%%%%%%%%%%%%%%%%%%%%%%%%%%%%%%%%%%%%%%%%%%%%%%%%%%%%%%%%%%%%%%%%%%%%%

  Fig.\ \ref{currents} shows typical current distributions obtained in the CRMF 
calculations for selected normal-, super- and hyperdeformed nuclei. Despite the fact 
that the moments of inertia of the SD and HD configurations are very close to the 
rigid-body values (Sec.\ \ref{pndefdep-mom}),  the presence of strong vortices\footnote{The 
existence of vortices at these points implies non-vanishing current 
circulations which are defined as 
$\bff{C}(\bff{r})=\bff{\nabla} \times \bff{j}(\bff{r})$ \cite{GR.78}.} 
demonstrates the dramatic deviation of the currents from rigid rotation. 
For example, the HD 
configurations in $^{92}$Mo (Fig.\ \ref{currents}a) and $^{108}$Cd (Fig.\ \ref{currents}b) 
show two strong vortices centered at $z\approx \pm 2$ fm. Note that the vortices (i.e. 
the curl) of the current fields are aligned or antialigned along a principal $x$-axis of 
the ellipsoid because of the use of one-dimensional cranking approximation. On the other 
hand, the HD configuration in $^{118}$Te (Fig.\ \ref{currents}c) shows one very strong 
vortice centered at $z=0$ fm, and 2 weaker vortices centered at $z \approx \pm 4.5$ fm. 
All three vortices rotate clockwise. The currents in the rotating frame of reference that 
is fixed to the body are caused by quantized motion of the fermions. Thus, the differences 
between the currents in $^{92}$Mo and $^{108}$Cd on one hand and the ones in $^{118}$Te on 
the other hand are caused by the differences in the underlying single-particle 
configurations. Contrary to the HD configurations, the current distributions in the SD 
configurations of $^{142}$Sm, $^{148}$Gd and $^{152}$Dy are characterized by a single very 
strong central vortice (Figs.\ \ref{currents}d-f). Current patterns in normal deformed
nuclei $^{100}$Sn and $^{118}$Ba look more disordered than in the SD and HD nuclei 
(Figs.\ \ref{currents}g,h). This is because three (four) large vortices in $^{100}$Sn 
($^{118}$Ba) are spread out over the volume of the nucleus. On the other hand, the current 
pattern is dominated by a single large central vortice in the ND configuration of $^{136}$Nd 
(Fig.\ \ref{currents}i).

  Note that all considered configurations are characterized by the weak current in the 
surface area. On the contrary, the average intrinsic current flows mainly in the nuclear 
surface in the semiclassical description of currents in normal and superfluid rotating 
nuclei \cite{DSK.85}. This underlines the importance of quantum mechanical treatment 
of the currents.

  The total current is the sum of Coriolis induced and magnetic potential induced 
currents. Total current is dominated by the Coriolis induced currents; magnetic potential 
induced currents represent approximately 5-20\% [30\%] of total current in the HD and SD 
[ND] nuclei shown in Fig.\ \ref{currents}. The only exception is $^{92}$Mo, in central 
region of which magnetic potential induced currents are larger than Coriolis induced 
currents by a factor close to 2.   The spatial distribution of Coriolis induced and 
magnetic potential induced currents is similar in the majority of nuclei shown in Fig.\ 
\ref{currents}. However, the spatial distribution of these two types of currents differ 
substantially in $^{92}$Mo, $^{146}$Gd and $^{118}$Ba.

  Comparing current patterns shown in Fig.\ \ref{currents}, one can conclude that for a 
system of non-interacting fermions, the total current, 
being the sum of the single-particle currents (see Eq.\ (\ref{current-eq})), is, in general, 
quite complicated.  This is a consequence of the fact that the localization, the strength 
and the structure of the current vortices created by a particle in a specific single-particle 
state depend on its nodal structure (see Ref.\ \cite{GR.78} and Sec.\ IIIC in Ref.\ 
\cite{AA.10}). In this respect it is important to mention the results of Ref.\ \cite{GR.78} 
which showed that Coriolis induced current for a single-particle in a slowly rotating anisotropic 
harmonic 
oscillator potential has, in fact, a rather simple structure. It exhibits a number of localized 
circulations with precisely predictable centers and sense of rotation. The centers of the 
circulations are found at the nodes and peaks of the oscillator eigenfunctions, thus, forming 
a rectangular array somewhat similar to a crystal lattice. 

  The wavefunction of the CRMF approach is more complicated than that of a rotating anisotropic 
harmonic oscillator because of the presence of spin-orbit interaction and the split of the 
wavefunction into large and small components. Moreover, there are magnetic potential induced
currents in addition to Coriolis induced ones. However, the analysis of single-particle 
vortices in rotating nuclei in general confirms the observations made in Ref.\ \cite{GR.78}. 
The typical features of the single-particle currents in the CRMF approach are considered below 
on the example of three neutron single-particle states occupied in the yrast SD configuration 
of $^{152}$Dy. The current and density distributions of these states are shown in Fig.\ \ref{sp-currents}. 
Let us first consider the $\nu [642]5/2^+$ state. The comparison of Figs.\ \ref{sp-currents}c and 
\ref{sp-currents}d  reveals that the rotation of a nucleus considerably increases the currents in 
this state. On the other hand, the density distribution is almost unaffected by rotation. The 
rotation of a nucleus also leads to a change in the structure of the circulations. At $\Omega_x=0.0$ MeV, 
there are three weak circulations centered around the nodes at 
$(y=0\,\, {\rm fm},z=0\,\, {\rm fm})$ and $(y=0\,\, {\rm fm},z\approx \pm 4 \,\,{\rm fm})$;
they are due to magnetic potential. Only two much stronger circulations are visible 
at $\Omega_x=0.5$ MeV: they are centered around the nodes located at  $(y=0\,\,{\rm fm}, 
z\approx \pm 2.5\,\,{\rm fm})$. This change of the structure of vortices can be attributed to
additional currents produced by the Coriolis term as well as to the change of the structure of 
wave function with increasing rotational frequency. The wave function in terms of two largest 
components has the 86\%[642]5/2+5\%[633]5/2 and 63\%[642]5/2+13\%[651]3/2 
structure\footnote{The percentages show the weights of respective components of the wave function 
in the total structure of the wave function. Note that only two largest components of the wave 
function are displayed.} at $\Omega_x=0.0$ MeV and $\Omega_x=0.5$ MeV, respectively.

  Even much large changes are induced by rotation into the structure of the $\nu [411]1/2^+$ state.
The wave function in terms of two largest components has the 57\%[411]1/2+23\%[651]1/2 and 
84\%[411]1/2+13\%[411]3/2 structure at $\Omega_x=0.0$ MeV and $\Omega_x=0.5$ MeV, respectively. One 
can see that the $\Delta N=2$ interaction, leading to a considerable admixture of the [651]1/2 
component into the structure of wave function, plays very important role at no rotation. The 
change in the wave function induced by rotation leads to a considerable changes both in the nodal 
structure of density distribution and in the current distribution (compare Fig.\ \ref{sp-currents}e 
with Fig.\ \ref{sp-currents}f).

   The wave function of the $\nu [770]1/2^-$ state is changed considerably by the rotation: its
structure in terms of two largest components is 62\%[770]1/2+17\%[761]1/2 at $\Omega_x=0.0$ MeV and 
39\%[770]1/2+28\%[761]3/2  and $\Omega_x=0.5$ MeV. The increase of rotational frequency does not lead 
to appreciable modifications in the density distribution but considerably decreases the strength of the 
currents and changes the shape of the circulations (see Figs.\ \ref{sp-currents}a,b). The latter is
a consequence of additional Coriolis induced currents. It is interesting that for this state the 
currents show maximum strength at the densities far below the maximum densities. This most likely 
explains relative weakness of the currents in this state as compared with those in the $\nu [411]1/2^+$ 
state. On the contrary, for many single-particle states the strongest currents are seen at or close to 
local increases in the densities (see Figs.\ \ref{sp-currents}c,d,f,g in the current manuscript and 
Fig.\ 8 in Ref.\ \cite{AA.10}).

   Our calculations show that the moments of inertia of the SD and HD configurations 
are very close to rigid-body values (Sec.\ \ref{pndefdep-mom}). However, the intrinsic 
currents show the dramatic deviations from rigid rotation. Usually the deviations 
from the rigid-body moment of inertia imply that the flow pattern must substantially deviate 
from  the current of a rigidly rotating mass distribution, i.e. there are strong net currents 
in the body-fixed frame \cite{DFPCU.04}. However, the opposite is not true: the closeness of 
the moments of inertia to rigid body value does not necessary implies that the current 
distribution should correspond to rigid rotation. On a microscopic level, the building 
blocks of the total current, namely, the single-particle currents certainly do not have a 
rigid-flow character; on the contrary, they have the vortex-flow character (see Fig.\ 
\ref{sp-currents}).

  Earlier non-relativistic studies also point to above discussed relations between currents
and rigid body moments of inertia. For example, it was shown in Ref.\ \cite{GR.78-0} for 
Schr{\"o}dinger equation that single-valuedness requirement for the wavefunction implies
non-existence of rigid-flow in a quantum fluid. Furthermore, it was demonstrated for a system 
of independent particles employing cranked harmonic oscillator potential that the current 
is not of the rigid-flow type even when the moment of inertia assumes the rigid-body value 
(\cite{GR.78-0}, see also Ref.\ \cite{R.76}).

  Current distributions shown in Figs.\ \ref{currents} and \ref{sp-currents} are typical for 
collective rotation around the $x-$axis perpendicular to the symmetry axis.  Note that the alignment 
of the  angular momentum vector of a particle is specified along the $x$-axis in one-dimensional 
cranking approximation (see also discussion in Sec.\ IIIC of Ref.\ \cite{AA.10}). The $\Omega=1/2$ 
orbitals are aligned with the axis of rotation ($x$-axis) already at no rotation. As a result, the 
single-particle angular momentum vector of the $\Omega=1/2$ orbitals performs the precession around 
the $x$-axis, thus orienting the currents predominantly in the $y-z$ plane. In addition, the 
$\Omega=1/2$ orbitals show vortices which are concentrated in the central region of nucleus. For 
the configurations with $\Omega \neq 1/2$, this mechanism of alignment becomes active only when the 
rotation sets up. Moreover, with increasing $\Omega$, the densities and currents are pushed away 
from the axis of symmetry of the nucleus toward the surface area (Figs.\ \ref{sp-currents}c,d and 
Fig.\ 8 in Ref.\ \cite{AA.10}).

%%%%%%%%%%%%%%%%%%%%%%%%%%%%%%%%%%%%%%%%%%%%%%%%%%%%%%%%%%%%%%%%%%%%%
\section{Frequency and configuration dependences of the impact 
of nuclear magnetism on the moments of inertia}
\label{Freq-dep}
%%%%%%%%%%%%%%%%%%%%%%%%%%%%%%%%%%%%%%%%%%%%%%%%%%%%%%%%%%%%%%%%%%%%%

   In this section, the frequency dependence of the impact of NM on 
the moments of inertia is studied using considerable number of SD and 
highly-deformed configurations in $^{60}$Zn obtained in unpaired CRMF
calculations. The properties of yrast SD band in this nucleus were 
well described in this formalism above band crossing which takes place 
at $\Omega_x \sim 1$ MeV \cite{Zn60,ARR.99}, while the CRHB+LN formalism 
gave good description of this band in the band crossing region \cite{Pingst-A30-60}. 
The neutron routhian diagram for this configuration obtained in the calculations 
with the NLSH parametrization is shown in Fig.\ 1 of Ref.\ \cite{ARR.99}; the 
results with the NL1 parametrization are similar to the ones obtained with NLSH. 
All proton and neutron states below the $Z=30$ and $N=30$ SD shell gaps are 
occupied in this configuration (note that proton routhian diagram is similar 
to the neutron one). The configurations are 
labelled by the shorthand notation $[n,p]$, where $n$ ($p$) is the number 
of occupied $g_{9/2}$ neutrons (protons). In this notation, the yrast SD 
band has the [2,2] configuration. Excited configurations under 
consideration are built by means of proton or/and neutron particle-hole 
excitations across  the $Z=30$ and $N=30$ SD shell gaps.

%%%%%%%%%%%%%%%%%%%%%%%%%%%%%%%%%%%%%%%%%%%%%%%%%%%%%%%%%%%%%%%%
\begin{figure}
\centering
\includegraphics[width=8.0cm]{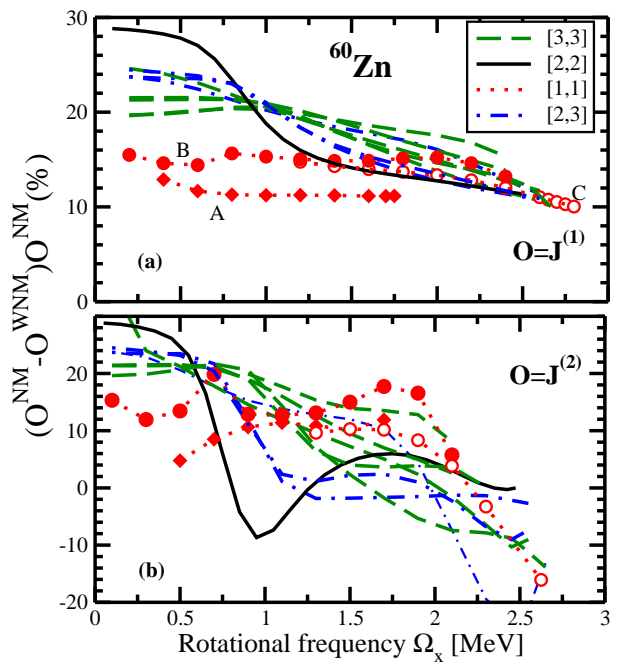}
\caption{(Color online) The contributions of NM to the dynamic ($J^{(2)}$) 
(panel (b)) and kinematic ($J^{(1)}$) (panel (a)) moments of inertia as a function 
of rotational frequency. Different color/line types are used for different 
groups of configurations characterized by the occupation of $g_{9/2}$ protons
and neutrons.
\label{fig-mom-contr-zn60}}
\end{figure}
%%%%%%%%%%%%%%%%%%%%%%%%%%%%%%%%%%%%%%%%%%%%%%%%%%%%%%%%%%%%%%%%

%%%%%%%%%%%%%%%%%%%%%%%%%%%%%%%%%%%%%%%%%%%%%%%%%%%%%%%%%%%%%%%%
\begin{figure*}
\centering
\includegraphics[width=14.0cm]{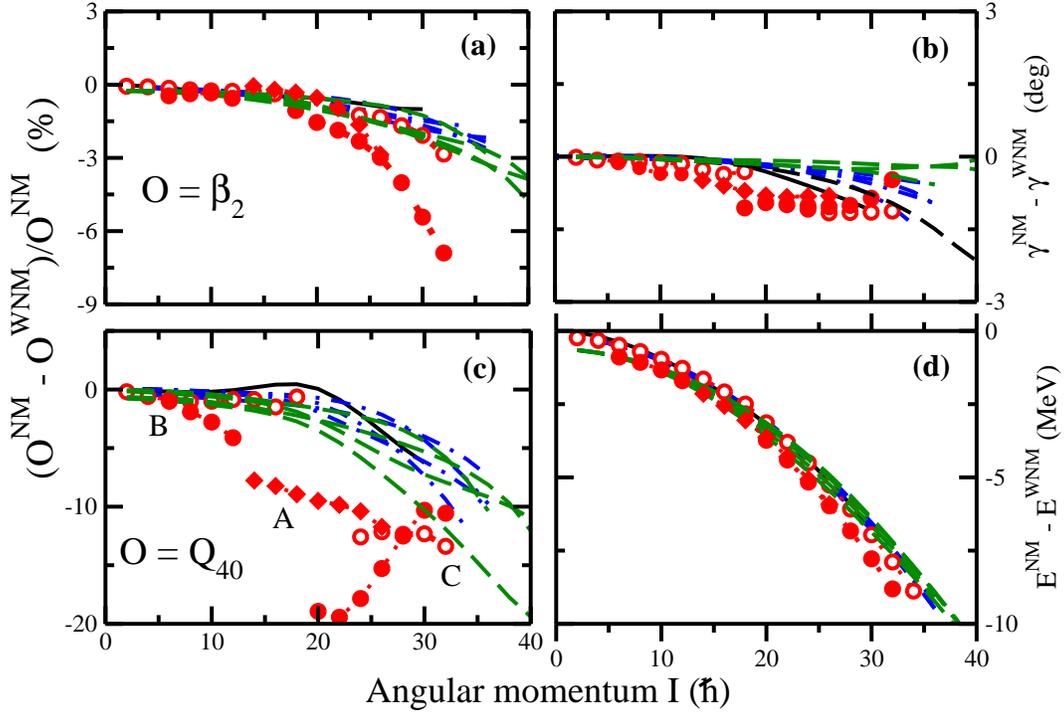}
\vspace{0.5cm}
\caption{(Color online) The contributions of NM to $\beta_2$-deformation (panel (a)),
$\gamma$-deformation (panel (b)), mass hexadecapole moment $Q_{40}$ and total energy
$E$ (panel (d)) as a function of angular momentum $I$. Note that the results on panels
(b) and (d) are shown in absolute values.
\label{zn60-res-def-en}}
\end{figure*}
%%%%%%%%%%%%%%%%%%%%%%%%%%%%%%%%%%%%%%%%%%%%%%%%%%%%%%%%%%%%%%%%

  The results of calculations for contributions of NM into dynamic ($\Delta J^{(2)}_{NM-contr}$) 
and kinematic ($\Delta J^{(1)}_{NM-contr}$) moments of inertia are shown in Fig.\ 
\ref{fig-mom-contr-zn60}. At low frequencies, the average contribution of NM into kinematic 
moment of inertia is slightly larger than 20\% (Fig.\ \ref{fig-mom-contr-zn60}a) and the
$\Delta J^{(1)}_{NM-contr}$ quantities show considerable dependence on configuration. The origin 
of the latter observation can be traced back to the specific features of some occupied 
single-particle orbitals. Let us consider as an example the [2,2] configuration. At low 
frequencies, the $\Delta J^{(1)}_{NM-contr}$ values for this configuration are considerable 
higher than the $\Delta J^{(1)}_{NM-contr}$ values averaged over all calculated configurations. 
This is due to the fact that upsloping branches of the proton and neutron $[440]1/2^+$ 
orbitals (in the $\Omega_x=0.0-0.7$ MeV range, see Fig.\ 1 in Ref.\ \cite{ARR.99}), 
characterized by the expectation values of the single-particle angular momentum
$\langle \hat{j}_{x}\rangle _{i}$ strongly affected by NM, are  occupied at $\Omega_x \leq 0.6$ MeV. 
At frequencies $\Omega_x \sim 0.8$ MeV, these orbitals strongly interact with proton and neutron 
$[431]3/2^+$ orbitals and exchange the character of the wavefunction. This leads to 
unpaired band crossing (see Ref.\ \cite{ARR.99}) which is seen in considerable changes of 
$\Delta J^{(1)}_{NM-contr}$ and $\Delta J^{(2)}_{NM-contr}$ quantities. The band crossing process is 
completed above $\Omega_x=1.1$ MeV, where the orbital labeled as $[440]1/2^+$ is downsloping as a 
function of rotational frequency (see Fig.\ 1 in Ref.\ \cite{ARR.99}). At these frequencies, the 
$\Delta J^{(1)}_{NM-contr}$ quantity for the [2,2] configuration is slightly below the value of 
$\Delta J^{(1)}_{NM-contr}$ averaged over all calculated configurations (Fig.\ \ref{fig-mom-contr-zn60}a).  
Note that this unpaired band crossing is not active in the [1,1] configurations because neither 
proton nor neutron $[440]1/2^+$ orbitals are occupied. The calculations also suggest that it is 
considerably suppressed in the [3,3] configurations due to the changes in the deformations and 
currents induced by the occupation of third $g_{9/2}$ orbital both in proton and neutron subsystems. 
However, the presence of this crossing is still visible (especially, in the $\Delta J^{(2)}_{NM-contr}$ 
quantity) in some [2,3] configurations.

   With increasing rotational frequency, the average contribution of NM into kinematic 
moments of inertia decreases and it falls below 15\% at $\Omega_x\sim 2.5$ MeV (Fig.\ 
\ref{fig-mom-contr-zn60}a). In addition, the configuration dependence of the 
$\Delta J^{(1)}_{NM-contr}$ quantities is weaker than the one at low frequencies. At these 
frequencies, the majority of occupied single-particle orbitals are either completely 
aligned or very close to complete alignment. However, NM do not modify the expectation 
values  of the single-particle angular momenta $\left< j_x \right>_i$  of completely 
aligned orbitals \cite{A.08}. As a result, only remaining orbitals, which are still 
aligning, contribute into $\Delta J^{(1)}_{NM-contr}$. The combined contribution of these 
orbitals into $\Delta J^{(1)}_{NM-contr}$ is smaller than the one at lower frequencies 
because the alignment of these orbitals is not far away from complete.

  The impact of NM on the dynamic moments of inertia is shown in Fig.\ \ref{fig-mom-contr-zn60}b
and it clearly displays much more complicated pattern as compared with the impact of NM
on the kinematic moments of inertia. The irregularities in the $\Delta J^{(2)}_{NM-contr}$ quantities 
are related to the band crossings. For example, the dip in the $\Delta J^{(2)}_{NM-contr}$ values 
of the $[2,2]$ configuration at $\Omega_x \sim 0.9$ MeV is caused by the unpaired band crossings 
which take place at different frequencies in the calculations with and
without NM. Similar deviations from smooth trend as a function of rotational frequency are 
visible in other configurations. However, one can see that for some configurations the 
contribution of NM into dynamic moment of inertia is a smooth function of rotational frequency 
over extended frequency range. In this frequency range, the configurations remain unchanged.
It is interesting that for some of these configurations the contributions of NM into dynamic 
moments of inertia are either close to zero or even negative; such features have not been seen
in the previous analyzes of the impact on time-odd mean fields on the dynamic moments of 
inertia \cite{KR.93,DD.95,AKR.96,AR.00}.

  The impact of NM on other physical observables of interest is shown in Fig.\ \ref{zn60-res-def-en},
in which the results of the NM and WNM calculations are compared as a function of total angular 
momentum. One can see that the quadrupole deformations $\beta_2$ (Fig.\ \ref{zn60-res-def-en}a) 
obtained  in the calculations with and without NM differ by less than 3\%. The only exception is 
configuration B for which this difference reaches 7\%. The difference in mass hexadecapole moments 
$Q_{40}$ obtained in the calculations with and without NM is larger but typically below 10\% (Fig.\ 
\ref{zn60-res-def-en}c); the only exception is the configuration B for which this difference
reaches 20\% at $I\sim 20\hbar$. The $\gamma$-deformations obtained in the calculations with and 
without NM 
differ by less than 1.5$^{\circ}$ (Fig.\ \ref{zn60-res-def-en}b). The only significant difference is 
seen in the total binding energies (Fig.\ \ref{zn60-res-def-en}d), where the NM solution is more bound 
than the WNM solution. This effect, which is due to the modifications in the moments of inertia 
induced by NM, is very large: additional binding due to NM reaches 7-8 MeV at spin 
$I=30\hbar$. These systematic results are consistent with the ones obtained in the previous 
studies of single SD configuration in $^{152}$Dy \cite{AR.00} and single terminating configuration
in $^{20}$Ne \cite{A.08}. They also give a hint why  the cranked models based on the phenomenological 
potentials like Woods-Saxon or Nilsson, which do not include time-odd mean fields \cite{DD.95}, are so 
successful in the description of experimental data. When considered as a function of spin the 
deformation properties of the rotating  system are only weakly affected by time-odd mean fields, 
and the proper renormalization of the moments of inertia \cite{AFLR.99} takes care of the $E$ versus 
angular momentum curve.

%%%%%%%%%%%%%%%%%%%%%%%%%%%%%%%%%%%%%%%%%%%%%%%%%%%%%%%%%%%%%%%
\section{Parametrization dependence of the contributions of NM
to the moments of inertia}
\label{Sec-par-dep}
%%%%%%%%%%%%%%%%%%%%%%%%%%%%%%%%%%%%%%%%%%%%%%%%%%%%%%%%%%%%%%%

%%%%%%%%%%%%%%%%%%%%%%%%%%%%%%%%%%%%%%%%%%%%%%%%%%%%%%%%%%%%%%%%%%%%%%%%%%%%%%%%%
\begin{figure*}
\centering
\includegraphics[width=14.0cm]{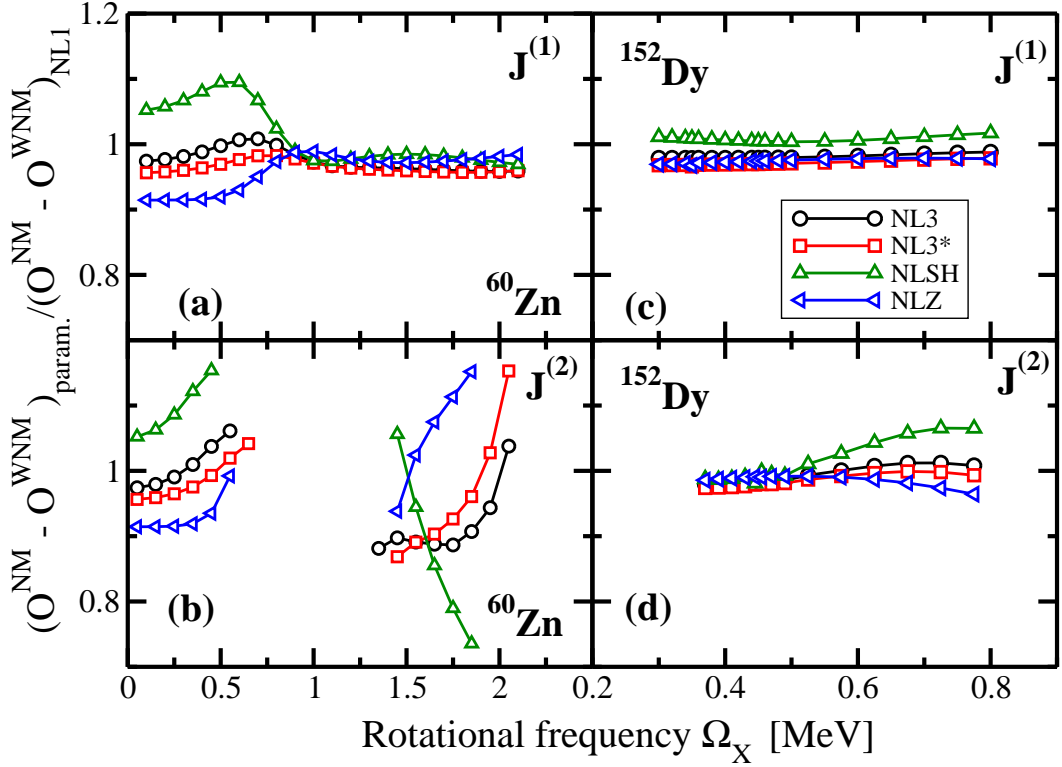}
\caption{(Color online) The contribution of NM to the physical observable $O$ 
for specific parametrization of the RMF Lagrangian (the $[O^{NM}-O^{WNM}]_{param}$ 
quantity) normalized to the one obtained in the NL1 parametrization. The results
of calculations are shown for yrast SD configurations in the $^{60}$Zn and $^{152}$Dy 
nuclei. Note that band crossing region is excluded in panel (b).
\label{Param}}
\end{figure*}
%%%%%%%%%%%%%%%%%%%%%%%%%%%%%%%%%%%%%%%%%%%%%%%%%%%%%%%%%%%%%%%%%%%%%%%%%%%%%%%%

  It was shown in Ref.\ \cite{AA.10} that additional binding due to NM in one-particle 
states only weakly depends on the RMF parametrization; this is also seen in the analysis 
of terminating states in Ref.\ \cite{A.08}. In this context, it is important to 
understand how the contributions of NM to the kinematic and dynamic moment of inertia
depend on the RMF parametrization.

  The dependence of the dynamic moments of inertia on the RMF parametrization has 
earlier been analyzed on the example of the SD bands in $^{151}$Tb and $^{143}$Eu in Ref.\ 
\cite{ALR.98} 
and in $^{58}$Cu and $^{60}$Zn in Ref.\ \cite{ARR.99} employing the NL1, NL3 \cite{NL3} and 
NLSH \cite{NLSH} parametrizations of the RMF Lagrangian.  The latter study includes also 
the results of calculations for kinematic moments of inertia. Additional calculations for 
these nuclei have also been performed with NL3* \cite{NL3*} and NLZ \cite{NLZ} parametrizations 
for the current manuscript. As follows from these results, the kinematic and dynamic moments 
of inertia depend only weakly on the parametrization of the RMF Lagrangian. Indeed, at given
frequency all the results for kinematic [dynamic] moments of inertia fit into the window 
which have a width equal to approximately 5\% ($\approx 8$\%) [approximately 6\% ($\approx 10$\%)]
of the value of kinematic [dynamic] moment of inertia  in the $A\sim 150$ ($A\sim 60$) region 
of superdeformation. The larger spread of calculated values 
in the $A\sim 60$ mass region are most likely due to (i) larger softness of potential energy 
surfaces in these nuclei as compared with the ones in the $A\sim 150$ region of superdeformation 
and (ii) to larger relative importance of each particle and, thus, model uncertainties 
in the description of their single-particle energies.

  Fig.\ \ref{Param} shows the dependence of the contributions of NM to the kinematic and dynamic 
moments of inertia on the parametrization of the RMF Lagrangian. For simplicity of comparison, 
these quantities are normalized to those obtained in the calculations with the NL1 parametrization. 
Very weak dependence (within 5\% window with respect of the NL1 results) of the contribution of 
NM to the kinematic moment of inertia on the RMF parametrization is seen in whole frequency range 
in $^{152}$Dy (Fig.\ \ref{Param}c) and at frequencies $\Omega_x \geq 0.75$ MeV in $^{60}$Zn 
(Fig.\ \ref{Param}a). In the latter nucleus the deviation from the NL1 results reaches 10\% at 
lower frequencies. The possible reasons for the larger dependence of calculated quantities
on the parametrization in $^{60}$Zn has been discussed above. On the other hand, the deviations 
from the NL1 results are larger for the dynamic moments of inertia. These deviations can be as large 
as 8\% at highest frequencies in the yrast SD configuration of $^{152}$Dy  (Fig.\ \ref{Param}d) and 
as large as 20\% in the yrast SD configuration in $^{60}$Zn (Fig.\ \ref{Param}b). Considering 
that the dynamic moment of inertia is related to the second derivative of the total energy with 
respect of spin, a larger dependence of the dynamic moment of inertia on the parametrization is 
expected.

  These values can be used to estimate the uncertainty in the definition of the moments of inertia
in the CRMF calculations due to the uncertainty in NM. The latter is related to the dependence of the  
$[O^{NM}-O^{WNM}]_{param}$ quantities (Fig.\ \ref{Param}) on the RMF parametrization discussed above.
 Dependent on  
nuclear system and configuration, the NM contribution to the total kinematic and dynamic moments of 
inertia is approximately 10-25\% (Secs.\ \ref{pndefdep-mom} and \ref{Freq-dep}).  Thus, the uncertainty 
of the definition of the absolute value of the total dynamic and kinematic moments of inertia due to 
the uncertainty in the definition of  NM is modest, being in range 0.5-5.0\%. The fact that the moments 
of inertia of rotational bands of different structure in unpaired regime are well (typically within 
5\% of experimental data \cite{AKR.96,ALR.98,ARR.99,VALR.05,AF.05,Dy154}) described in the CRMF 
calculations strongly suggests that NM and its impact on the moments of inertia is reasonably well 
defined in the CDFT theory.

%%%%%%%%%%%%%%%%%%%%%%%%%%%%%%%%%%%%%%%%%%%%%%%%%%%%
\section{Terminating states}
\label{Sec-term-states}
%%%%%%%%%%%%%%%%%%%%%%%%%%%%%%%%%%%%%%%%%%%%%%%%%%%%

%%%%%%%%%%%%%%%%%%%%%%%%%%%%%%%%%%%%%%%%%%%%%
\begin{figure*}[ht]
\centering
\includegraphics[width=16.0cm]{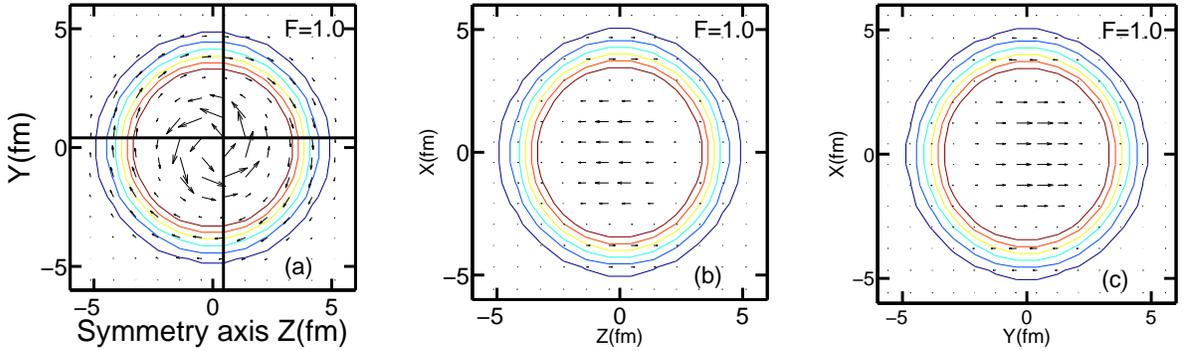}
\caption{ Neutron current distributions  {\bff j}$^n$({\bff r}) in the intrinsic frame in 
the $y-z$ plane (at $x=0.416$ fm) (left panel), in the $x-z$ plane (at $y=0.416$ fm) (middle 
panel), and in the $x-y$ plane (at $z=0.457$ fm). These distributions are shown for the state  
of $^{47}$V  terminating at $I=17.5^+$. The currents are plotted at arbitrary units for better 
visualization. The shape and size of the nucleus are indicated by density lines which are 
plotted in the range $0.01-0.06$ fm$^{-3}$ in step of 0.01 
fm$^{-3}$. Vertical and horizontal lines on left panel show the cross-sections at which 
the currents in the $z-x$ plane (middle  panel) and in the $y-x$ plane (right panel) 
are plotted, respectively.}
\label{Term-curr-distr}
\end{figure*}
%%%%%%%%%%%%%%%%%%%%%%%%%%%%%%%%%%%%%%%%%%%

%%%%%%%%%%%%%%%%%%%%%%%%%%%%%%%%%%%%%%%%%%%%%
\begin{figure*}[ht]
\centering
\includegraphics[width=16.0cm]{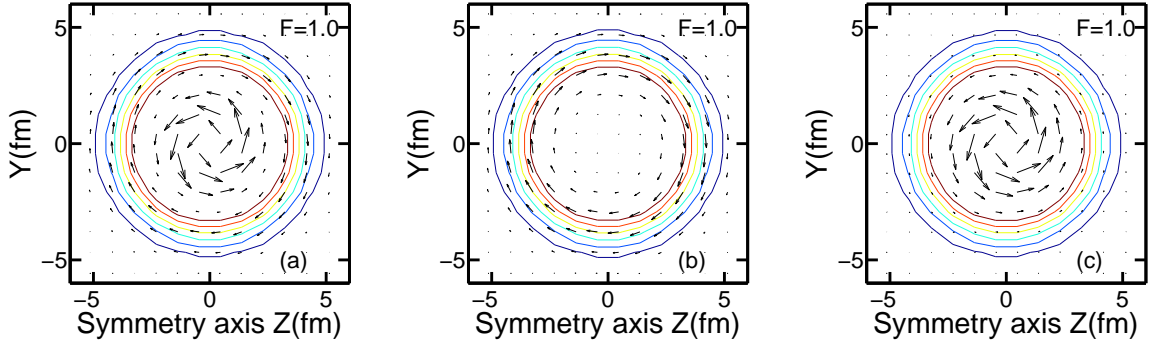}
\caption{    Neutron current distributions  {\bff j}$^n$({\bff r}) in the intrinsic frame 
in the $y-z$ plane (at $x=0.407$ fm). These distributions are shown for the state of 
$^{47}$V  terminating at $I=17.5^+$. Left, middle and right panels show total, Coriolis 
induced and magnetic potential induced currents, respectively. See caption of Fig.\ 
\ref{Term-curr-distr} for other details. }
\label{Term-curr-distr-V47}
\end{figure*}
%%%%%%%%%%%%%%%%%%%%%%%%%%%%%%%%%%%%%%%%%%%

  The majority of rotational bands which do not have large deformation at spin zero
will terminate in a non-collective terminating state at $I_{max}$ \cite{CHO,Ragbook,AFLR.99} 
\footnote{Only recently the evidences for non-termination of some rotational bands 
at $I_{max}$ have been found \cite{Kr74-no-term}.}. The regime of nuclear motion in
terminating state is usually referred as 'non-collective rotation' \cite{S.book,AFLR.99}. 
This is because of the fact that for an axially symmetric potential, the nucleon orbitals are not 
influenced by the rotation around the symmetry axis of this potential; thus, collective 
rotation about that axis is not possible. Non-collective rotation is also realized in the 
aligned states such as `yrast traps' (or 'yrast isomers') \cite{And78,RS.80,VDS.83}. The 
study of terminating states in the context of understanding of time-odd  mean fields is 
of considerable interest because of several reasons. First, time-odd mean fields provide an 
additional binding to the energies of the specific configuration, and this additional binding 
increases with spin and has its maximum exactly at the terminating state \cite{A.08}. This 
suggests that the terminating states can be an interesting probe of time-odd mean fields 
\cite{ZSW.05,SW.05,A.08} {\it provided that other effects can be reliably isolated 
\cite{A.08}.}  Second, at the band termination, the NM does not modify either total angular 
momentum or the expectation values of the single-particle angular momenta 
$\langle \hat{j}_x \rangle_i$ \cite{A.08}.  Third, terminating state is a 
(multi)-particle+(multi)-hole non-collective state in which the 
angular momenta of all particles and holes outside the core are aligned along the symmetry 
axis.

%{\bf   Calculation of terminating states within density functional theories is also 
%of considerable interest since it enables the study of time-odd mean fields associated 
%with currents and spin densities \cite{SW.05}; note that contrary to Skyrme EDF theory
%spin densities do not affect explicitly binding energies in CDFT (see Eqs.\ 
%\ref{Etot}-\ref{E-cm} below).}

 We will consider in this section the $\pi (d_{3/2})^{-1}_{1.5}(f_{7/2})^4_{8} \otimes \nu (f_{7/2})^4_{8}$ 
terminating state in $^{47}$V, which has $I_{max}=17.5^+$, as an example. The structure of this state is 
given with respect of the $^{40}$Ca core. This state is characterized by the largest impact of NM on 
the binding  energies amongst terminating states studied in Ref.\ \cite{A.08}. It is nearly spherical with the 
quadrupole deformation $\beta_2\sim 0.03$ (Fig.\ 6 in Ref.\ \cite{A.08}). Our goal is to understand 
the impact of NM on the current distribution and microscopic origin of additional binding due to NM.

  In terminating states, the angular momenta of valence particles and holes are aligned along 
the symmetry axis ($x$-axis). As a consequence they perform precession around this axis, 
generating azimuthal currents with respect to the symmetry axis. This is illustrated in Fig.\ 
\ref{Term-curr-distr}.  One can see two azimuthal circulations in the $y-z$ plane: the circulation 
in the central region of nucleus is directed counterclockwise while the one in the surface region 
is directed clockwise. Fig.\ \ref{Term-curr-distr-V47} shows total [left panel], Coriolis induced 
[middle panel] and magnetic potential induced [right panel] currents. One can see that surface
circulation is generated by the Coriolis term, while the central circulation by the magnetic
potential. The currents in the $x-z$ and  $x-y$ planes are perpendicular to the $x$-axis (Fig.\ 
\ref{Term-curr-distr}). This clearly shows that the currents are azimuthal.

  In Ref.\ \cite{AA.10}, the polarization effects induced by NM have been investigated in 
one- and two-particle configurations of odd and odd-odd non-rotating nuclei.  Terminating 
states differ significantly from these configurations. First, they are multi-particle+multi-hole 
configurations. For example, the $\pi (d_{3/2})^{-1}_{1.5}(f_{7/2})^4_{8} \otimes \nu (f_{7/2})^4_{8}$ 
terminating state in $^{47}$V has 8 particles and 1 hole outside the $^{40}$Ca core. Second, 
the alignment of the angular momenta of these particles and holes generates considerable 
total angular momentum ($I=17.5\hbar$ in the discussed terminating state of $^{47}$V) aligned 
along the axis of symmetry; this momentum is much larger than the one in odd and odd-odd 
nuclei studied in Ref.\ \cite{AA.10}. Third, terminating states are characterized by the 
azimuthal currents with respect to the symmetry axis, while the states in non-rotating 
nuclei are characterized by the currents shown in Figs.\ 7-9 of Ref.\ \cite{AA.10}. Thus, 
it is interesting to see how these differences affect the polarization effects induced by 
NM and whether these polarization effects are similar in nature for these two classes of 
non-collective states, namely, low-spin one- and two-particle configurations of non-rotating 
nuclei and high-spin terminating states.

  In order to facilitate the discussion, we split the total energy of the system 
(Refs.\ \cite{KR.89,AKR.96}) into different terms as\footnote{We follow Refs.\ 
\cite{RGL.97,AA.10} in the selection of the signs of the energy terms.}
\begin{eqnarray}
E_{tot} & = & E_{part} + E_{Cor} + E_{cm} - E_{\sigma} - E_{\sigma NL} - E_{\omega}^{TL}  \nonumber \\
       &   &  - E^{TL}_{\rho} - E_{\omega}^{SL} - E^{SL}_{\rho} - E_{Coul},
\label{Etot}
\end{eqnarray}
where $E_{part}$, $E_{Cor}$, and $E_{cm}$ represent the contributions from fermionic
degrees of freedom,  whereas the other terms are related to mesonic (bosonic) degrees 
of freedom.
In Eq.\ (\ref{Etot})
\begin{eqnarray}
E_{part}=\sum_i^A \varepsilon_i 
\label{Epart}
\end{eqnarray}
is the energy of the particles moving in the field created by the mesons ($\varepsilon_i$ 
is the energy of $i$-th particle and the sum runs over all occupied proton and neutron 
states),
\begin{eqnarray}
E_{Cor}= \Omega_x \sum_i^A \langle i | \hat{j}_x | i \rangle
\label{Ecor}
\end{eqnarray}
is the energy of the Coriolis term,
\begin{eqnarray}
E_{\sigma}=\frac{1}{2}\,\, g_{\sigma} \int d^3r\,\, \sigma({\bff r}) \left[\rho_s^p({\bff r})+\rho_s^n({\bff r})\right]
\end{eqnarray}
is the linear contribution to the energy of isoscalar-scalar $\sigma$-field,
\begin{eqnarray}
E_{\sigma NL} = \frac{1}{2} \int d^3r \left[ \frac{1}{3}\,g_2\, \sigma^3({\bff r}) + \frac{1}{2}\, g_3\, \sigma^4({\bff r}) 
\right]
\end{eqnarray}
is the non-linear contribution to the energy of isoscalar-scalar $\sigma$-field,
\begin{eqnarray}
E^{TL}_{\omega} = \frac{1}{2}\,\, g_\omega \int d^3r\,  \omega_0({\bff r}) \left[ \rho_v^p({\bff r})+\rho_v^n({\bff r}) \right]
\end{eqnarray}
is the energy of the time-like component of isoscalar-vector $\omega$-field,
\begin{eqnarray}
E^{TL}_{\rho} = \frac{1}{2} \,\,g_\rho \int d^3r  \rho_0({\bff r}) \left[\rho_v^n({\bff r})-\rho_v^p({\bff r})\right]
\end{eqnarray}
is the energy of the time-like component of isovector-vector $\rho$-field,
\begin{eqnarray}
E^{SL}_{\omega} = - \frac{1}{2}\,\,  g_\omega \int d^3r\, {\bff \omega}({\bff r}) \left[ {\bff j}^p({\bff r})+{\bff j}^n({\bff r}) 
\right]
\label{E-s-omega}
\end{eqnarray}
is the energy of the space-like component of isoscalar-vector $\omega$-field,
\begin{eqnarray}
E^{SL}_{\rho} = - \frac{1}{2}\,\, g_\rho \int d^3r\,  {\bff \rho}({\bff r}) \left[ {\bff j}^n({\bff r})-{\bff j}^p({\bff r}) 
\right]
\label{rho-space}
\end{eqnarray}
is the energy of the space-like component of isovector-vector $\rho$-field,
\begin{eqnarray}
E_{Coul} = \frac{1}{2}\,\,  e \int d^3r  A_0({\bff r}) \rho^p_v({\bff r})
\end{eqnarray}
is the Coulomb energy, and
\begin{eqnarray}
E_{cm}=-\frac{3}{4}\hbar \omega_0=-\frac{3}{4}\,\, 41 A^{-1/3}\,\, {\rm MeV}
\label{E-cm}
\end{eqnarray}
is the correction for the spurious center-of-mass motion approximated by its value in
a non-relativistic harmonic oscillator potential.

%%%%%%%%%%%%%%%%%%%%%%%%%%%%%%%%%%%%%%%%%%%%%%%%%%%%%%%%%%%%%%%%%%%%%%%%%%%%%%%%%%%%%%%%%%%%
\begin{table}[h]
\caption{Impact of NM on different terms of the total energy (Eq.\ (\protect\ref{Etot})) 
in the state of $^{47}$V terminating at $I=17.5^+$. Column (2) shows the absolute 
energies [in MeV] of different energy terms in the case when NM is neglected. Columns 
(3) and (4) list the changes $\Delta E_{i}=E^{NM}_{i}-E^{WNM}_{i}$ [in MeV] in the energies 
of these terms induced by NM in self-consistent [column (3)] and perturbative [column (4)] 
calculations. Note that only nonzero quantities are listed in column (4). The results of
calculations are obtained at rotational frequency $\Omega_x=2.4$ MeV. The energies  of the 
Coriolis term $E_{Cor}$ and particle energy $E_{part}$ depend on frequency. The latter 
takes place through the modifications of the energies of single-particle states with 
frequency (see, for example, Sec. 3.8 and Fig. 6a in Ref.\ \cite{AFLR.99} for more details 
on a construction of terminating state). However, the  sum $E_{Cor}+E_{part}$ do not depend 
on frequency. Other remaining quantities shown in column (1) and the $\Delta E_i$ quantities 
shown in columns (3) and (4) are frequency independent.}
\label{NM-term}
\vspace{0.5cm}
\begin{center}
\begin{tabular}{|c|c|c|c|c|} \hline
Quantity            &  $E_i^{WNM}$  & $\Delta E_i$ & $\Delta E_i^{pert}$ \\ \hline
     1              &      2       &       3            &      4       \\ \hline
$E_{part}$          & $-1217.668$  &  $-15.5$      & -7.296             \\
$E_{Cor}$           &  42.0        &    0          &                    \\
$E_{\sigma}$        &  $-6381.875$  &  $-85.231$    &     \\          
$E_{\sigma NL}$     &   109.368  &  $-2.815$        &     \\        
$E_{\omega}^{TL}$      &  5371.721 &  $79.291$       &       \\        
$E_{\omega}^{SL}$      &  0.0     &  $-3.51$          & -3.51   \\       
$E_{\rho}^{TL}$        &  0.549   &  $-0.002$        & \\                
$E_{\rho}^{SL}$        &  $ 0.0$     &  $-0.038$     & -0.038    \\      
$E_{Coul}$          &  $ 102.168$  &  $0.515$     & 0.240  \\         
$E_{cm}$            &  $ -8.521 $  &  0.0         &    \\           
$E_{tot}$           &  $ -386.121$  & $-3.704$   & -3.983  \\  \hline \hline
\end{tabular}
\end{center}
\end{table}
%%%%%%%%%%%%%%%%%%%%%%%%%%%%%%%%%%%%%%%%%%%%%%%%%%%%%%%%%%%%%%%%%%%%%%%%%%%%%%%

   Polarization effects induced by NM are investigated by considering NM impact 
on different terms of the total energy (Eq.\ (\protect\ref{Etot})). The results of this 
study are shown in Table \ref{NM-term}.  Similar to Ref.\ \cite{AA.10}, the $E_{\rho}^{TL}$  and 
$E_{\rho}^{SL}$ terms are only weakly influenced by NM, and thus, they will not be discussed 
in detail. Somewhat stronger impact of NM is seen in the $E_{Coul}$, $E_{\sigma NL}$, and 
$E_{\omega}^{SL}$ terms. Note that only last term was appreciably affected in low-spin 
configurations of odd and odd-odd nuclei in Ref.\ \cite{AA.10}. Much larger polarization 
effects are seen in the $E_{part}$, $E_{\sigma}$ and $E_{\omega}^{TL}$ terms. The $E_{\sigma}$ 
and $E_{\omega}^{TL}$ terms depend only indirectly on time-odd mean fields through the 
polarizations of time-even mean fields induced by NM \cite{AA.10}. One should keep in 
mind that only the $E_{\sigma}+E_{\omega}^{TL}$ quantity has a deep physical meaning, as 
it defines a nucleonic potential; this sum is modified by NM  on -5.9 MeV.

   Comparing these results with those presented in Ref.\ \cite{AA.10}, one can conclude 
that polarization effects for different total energy terms in terminating state under 
study are stronger by at least one order of magnitude than in low-spin one- and two-particle 
configurations of 
non-rotating nuclei. This is a consequence of the fact that all particles (8) and holes 
(1) outside the $^{40}$Ca core participate in building the total angular momentum and 
currents in terminating state, while only one (two) particle(s) participate in generating 
the currents in non-rotating odd (odd-odd) nuclei \cite{AA.10}. Despite that the 
relative impact of NM on different terms of the total energy is, in general, similar in 
these two classes of  non-collective states (compare Table I in the present manuscript 
with Tables I, II, and IV in Ref.\ \cite{AA.10}).

   Total modifications of the energies due to NM in the mesonic sector are -11.79 MeV. Only one 
third of these modifications comes from the terms ($E_{\omega}^{SL}$, $E_{\omega}^{SL}$) which 
directly depend on nucleonic currents, whereas the rest from the modifications of time-even 
mean fields induced by NM.

   It is interesting to compare the results of self-consistent and perturbative 
calculations\footnote{Fully self-consistent calculations with NM 
provide a starting point for perturbative calculations. Using their 
fields as input fields, only one iteration is performed in the calculations 
without NM: this provides perturbative results. Time-even mean fields are the same 
in both (fully self-consistent and perturbative) calculations. Then, the impact of 
time-odd mean fields on calculated quantities (for example, different terms in the 
total energy (Eq.\ (\ref{Etot})) is defined as the difference between 
the values of this quantity obtained in these two calculations. In this
way, the pure effects of time-odd mean fields in fermionic and mesonic 
channels of the model are isolated because no polarization effects are 
introduced into time-even mean fields \cite{AA.10}.}. The 
$\Delta E_{i}^{pert}=E^{NM}_{i}-E^{WNM}_{i}$ quantities will be used for simplicity in 
further discussion. These quantities are shown in columns 3 and 4 of Table \ref{NM-term}. 
The $\Delta E_{\sigma}$, $\Delta E_{\sigma NL}$,  $\Delta E_{\omega}^{TL}$ and $\Delta E_{\rho}^{TL}$ 
quantities are zero in perturbative calculations because time-even fields are fixed in these 
calculations. The 
$\Delta E_{\omega}^{SL}$ and $\Delta E_{\rho}^{SL}$ are the same in self-consistent 
(column 3) and perturbative (column 4) calculations because the $E_{\omega}^{TL}$ and 
$E_{\rho}^{TL}$ terms depend only on time-odd mean fields, which are the same in
the parts of the calculations that include NM. Particle energies $E_{part}$ are strongly 
modified by NM in self-consistent calculations; they change by $-15.5$ MeV. Perturbative calculations 
show that only one half of $\Delta E_{part}$ is coming directly from time-odd mean fields 
(see Sec. IVB of Ref.\ \cite{AA.10} for more details on this mechanism), the rest  is due 
to polarization effects in time-even fields induced by NM. The same is true for the Coulomb 
energy term $E_{Coul}$. 

  It is evident from Table \ref{NM-term} that
\begin{eqnarray} 
\Delta E_{tot}^{self-const} \approx \Delta E_{tot}^{pert}.
\label{Eqq}
\end{eqnarray}
Note that the superscript {\it 'self-const'} and {\it 'pert'} refers to fully self-consistent 
and perturbative results. The analysis of polarization effects in other terminating states of 
the $A\sim 40$ mass region shows the same relation. {\it These results clearly indicate that 
the additional binding due NM (the $E^{NM}-E^{WNM}$ quantity) is defined mainly by time-odd 
fields and that the polarization effects in fermionic and mesonic sectors of the model cancel 
each other to a large degree.} The same result has been earlier obtained in the analysis of 
non-rotating nuclei in Ref.\ \cite{AA.10}.

%%%%%%%%%%%%%%%%%%%%%%%%%%%%%%%%%%%%%%%%%%%%%%%%%%%%%%%%%%
\section{Signature-separated configurations}
\label{Sec-sign-sep}
%%%%%%%%%%%%%%%%%%%%%%%%%%%%%%%%%%%%%%%%%%%%%%%%%%%%%%%%%%

  Signature separation phenomenon induced by time-odd mean fields has 
been found earlier in excited 4-particle SD configurations of $^{32}$S 
\cite{MDD.00,Pingst-A30-60}, and very recently in 2-particle configurations
of nonrotating odd-odd nuclei in Ref.\ \cite{AA.10}. It reveals itself in a 
considerable energy splitting of the $r_{tot}=+1$ and $r_{tot}=-1$ branches of 
the configurations which have the same structure in terms of occupation of 
single-particle states with given Nilsson labels. Such a signature separation 
could not have been obtained in phenomenological cranking models, such as the 
ones using the Woods-Saxon or Nilsson potentials, since time-odd mean fields
are absent in these models.

However, the description of rotating $N\approx Z$ nuclei requires isospin 
projection  \cite{SDNR.10} which can modify above mentioned results. Since this 
projection is beyond the current framework, we concentrate at the nuclei away from 
the $N=Z$ line.  The analysis of Ref.\ \cite{AA.10} shows that signature separation 
is expected also in such nuclei, but it is weaker as compared with the one seen in 
the nuclei around the $N=Z$ line.  Unfortunately, the survey of odd-odd $A=20-52$ 
nuclei (some of which were studied in Ref.\ \cite{AA.10}) does not reveal experimental 
bands in the nuclei away from the $N=Z$ line in which signature separation is expected.

\begin{figure}
\centering
\includegraphics[width=8.0cm]{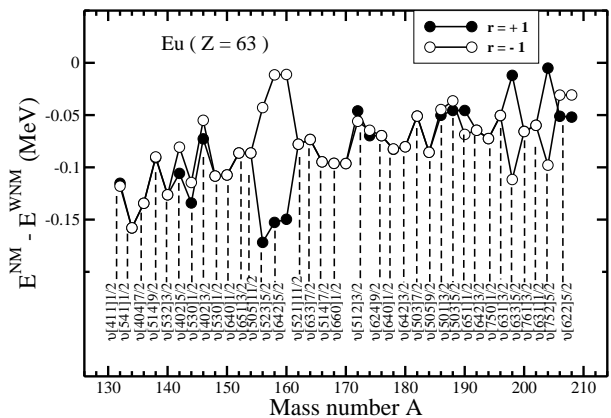}
\caption{The impact of NM on binding energies of the configurations in odd-odd Eu 
($Z=63$) nuclei at $\Omega_x=0.0$ MeV. The $E^{NM}-E^{WNM}$ quantity is shown for different 
total signatures $r$. Proton configuration has blocked $\pi [532]5/2^+$ state for all 
isotopes, while the neutron configuration changes as a function of mass number. 
Configurations are labeled with the Nilsson labels of blocked odd neutron state; 
configurations at and to the right of the Nilsson label up to the next Nilsson label 
have the same blocked neutron state.
\label{Eu-sep}}
\end{figure}

Fig.\ \ref{Eu-sep} shows that signature separation phenomenon can also be present in heavier 
nuclei. This figure shows the results of calculations for odd-odd Eu isotopes in which odd 
proton occupies fixed $\pi [532]5/2^+$ state, and odd neutron occupies different neutron states 
of the $r=\pm i$ signatures along the isotope chain. Additional binding due to NM (the 
$E^{NM}-E^{WNM}$ quantity, see Ref.\ \cite{AA.10} for more details) is shown for total proton-neutron 
configurations with different total signatures $r_{tot}=\pm i$. Significant signature separation 
(on the level of 100-150 keV) is seen in the $\pi [532]5/2 \otimes \nu [523]5/2$ ($^{156}$Eu), 
$\pi [532]5/2 \otimes \nu [642]5/2$ ($^{158,160}$Eu), $\pi [532]5/2 \otimes \nu [633]5/2$ ($^{196}$Eu), 
and $\pi [523]5/2 \otimes \nu [752]5/2$ ($^{204}$Eu) configurations. Either $r=-1$ or $r=+1$ states 
can be more bound in signature separated configurations of Eu isotopes (Fig.\ \ref{Eu-sep}). This 
depends on  mutual orientation of proton and neutron currents induced by odd proton and odd neutron; 
the state with the same orientation of these currents is more bound.

\begin{figure}
\centering
\includegraphics[width=8.0cm]{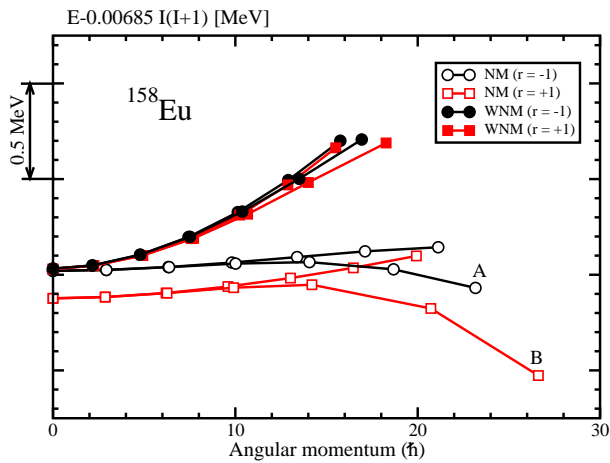}
\caption{(Color online)  The energies of calculated bands in $^{158}$Eu based 
on the $\pi [532]5/2^{\pm} \otimes \nu [642]5/2^{\pm}$  two-particle configurations as a 
function of angular momentum with rigid rotor reference subtracted. The results 
of the calculations with and without NM are shown. 
\label{Eu158-exp}}
\end{figure}

   Fig.\ \ref{Eu158-exp} illustrates that four rotational sequences (two with 
total signature $r_{tot}=+1$ and two with $r_{tot}=-1$) can be built in the 2-particle 
configurations $\pi |a>(r=\pm i)  \otimes \nu |b> (r=\pm i)$ (where $|a>$ and $|b>$ 
indicate the blocked proton and neutron Nilsson states, respectively) of odd-odd 
nuclei. For the case of $^{158}$Eu we consider 2-particle configurations based
on the $|a>\, = [532]5/2$ and $|b>\, = [642]5/2$ states. In the 
WNM calculations, the $r_{tot}=+1$ and $r_{tot}=-1$ configurations are almost degenerate 
in energy up to spin $I\sim 10\hbar$ (Fig.\ \ref{Eu158-exp}). On the contrary, there 
is a considerable signature separation ($\approx 150$ keV) due to time-odd mean fields between these 
configurations in the calculations with NM. {\it This feature is a strong spectroscopic 
fingerprint of the presence of time-odd mean fields.}  Note that rotational sequences
A and B undergo unpaired band crossings at $I\sim 20\hbar$.

  Unfortunately, the experimental data on  odd-odd Eu nuclei also do not reveal the 
configurations discussed above in which the signature separation is expected. This situation 
may be resolved by a systematic search of signature separated configurations both in the 
experimental data on odd-odd rare-earth nuclei and in the model calculations. The best way 
to confirm the existence of this phenomenon would be to find in model calculations the 
configurations of odd-odd nuclei which show no signature splitting in the absence of 
time-odd mean fields and measurable signature separation in the presence of time-odd mean 
fields, and then to find their experimental counterparts which show signature separation.
However, such an investigation is definitely beyond the scope of the current study. The 
difficulty of such a study is also underlined by the fact that existing interpretations of 
two-quasiparticle configurations in odd-odd nuclei are based on Woods-Saxon or Nilsson 
potentials. In these potentials, the signature degeneracy is considered to be a strong 
fingerprint of specific configurations. However, time-odd mean fields in EDF provide additional 
mechanism of breaking the signature degeneracy, so the experimental data on such configurations 
has to be reanalyzed in density functional calculations.

%%%%%%%%%%%%%%%%%%%%%%%%%%%%%%%%%%%%%%%%%%%%%%%%%%%%%%%%%%%%%%%%%%%%%%%%%%%%%%%%%
%%%%%%%%%%%%%%%%%%%%%%%%%%%%%%%%%%%%%%%%%%%%%%%%%%%%%%%%%%%%%%%%%%%%%%%%%%%%%%%%

  Although the model calculations clearly indicate the important role of time-odd
mean fields in creating signature separation phenomenon in odd-odd nuclei, the 
direct comparison with experiment will be complicated by the number of model 
limitations which are related to
\begin{itemize}   
\item       
the presence of residual proton-neutron interaction of unpaired proton and 
neutron,

\item
the coupling scheme of angular momenta vectors of unpaired proton and neutron at 
low spin.

\end{itemize}

  In odd-odd nuclei the angular momenta of unpaired proton and unpaired neutron in 
2-quasiparticle configurations can be coupled either in parallel or antiparallel
fashion, namely into $K_> = \Omega_p + \Omega_n$, and $K_< = |\Omega_p - \Omega_n|$, 
where $\Omega_{p(n)}$ represents the projection of single quasiparticle  angular 
momentum of proton (neutron) on the axis of symmetry. For example, in $^{158}$Eu
this will lead to rotational sequences with $K_<=0$  and $K_>=5$. The degeneracy 
of the bandheads of the $K_{><}$ doublet pair (called Gallagher-Moszkowski doublet 
\cite{GM.58}) is lifted by inclusion of the residual proton-neutron interaction and 
also by the zero-point rotational energy. Relative energy ordering of the $K_{>}$ 
and $K_{<}$ bands is determined by the empirical Gallagher-Moszkowski (GM) rule which 
places the spin-parallel band lower in energy than its spin-antiparallel counterpart 
\cite{GM.58} in odd-odd nuclei (and vise versa in even-even nuclei \cite{GJ.92}) and 
has only few exceptions \cite{BPO.76,NKSSN.94}. Another important consequence of the 
residual interaction of unpaired nucleons is the observed shift of the odd- and 
even-spin rotational levels relative to each other in the $K=0$ bands; this feature is 
generally referred to as the Newby or odd-even shift \cite{N.62}.

  Residual proton-neutron interaction of unpaired nucleons
 is neglected in the cranking models; we are 
not aware about any publication which includes it. So, neither Gallagher-Moszkowski
splittings nor Newby shifts can be described in the current calculations. It is also 
necessary to recognize that 2-quasiparticle configurations in
odd-odd and even-even nuclei show a daunting complexity due to the high
density of states and the large number of couplings and interactions possible.
The problem of the description of the Gallagher-Moszkowski splittings and Newby 
shifts is far from being settled even in the framework of conventional 
particle+rotor model \cite{BPO.76,NKSSN.94,JSHSet.98,JKSH.89,GJ.92}. For 
example, the  residual interaction of unpaired proton and neutron in odd-odd nuclei 
shows pronounced dependence on the mass region under study \cite{BPO.76,NKSSN.94}. 
It is even more difficult to understand why in 2-quasiparticle configurations of 
the rare-earth region different residual interactions are required to describe 
the interaction between unpaired proton and neutron in odd-odd nuclei and between 
unpaired protons (neutrons) in even-even nuclei \cite{GJ.92} despite the expectations 
that they should be the same due to isospin symmetry. To our knowledge, 
self-consistent description of Gallagher-Moszkowski splittings and Newby shifts 
has been attempted only in the framework of the rotor+two-quasiparticle model based
on Skyrme Hartree+Fock approach in Ref.\ \cite{BLMQ.87}.

  At zero rotational frequency the angular momenta of odd proton and odd 
neutron are aligned (parallel or anti-parallel) with the symmetry axis which 
leads to  band-head states with $K_> = \Omega_p + \Omega_n$ and 
$K_< = |\Omega_p - \Omega_n|$. However, in one-dimensional cranking approximation 
nuclear configuration on top of which rotational sequence is built does not 
depend on coupling of $\Omega_p$ and $\Omega_n$. This is well known (although 
seldom stressed) deficiency of one-dimensional cranking approximation.
However, with increasing rotational frequency the angular momenta of odd proton 
and odd neutron start to align with the axis of rotation which is perpendicular to the 
axis of symmetry. Although it is tempting to employ tilted axis cranking (TAC) 
approximation for the description  of the combination of these two angular momenta 
coupling schemes at low spin, this does not resolve the problem of the description 
of signature separation since signature is no longer good quantum number in the TAC 
approximation \cite{F.01}. On the contrary, one-dimensional cranking approximation used 
in the current manuscript has a clear advantage that it properly accounts for the 
alignments of valence particles and holes along the axis of rotation at medium and high 
spins where $I\geq K$ \cite{VDS.83}, and thus provides correct description of signature 
separation at these spins.

%%%%%%%%%%%%%%%%%%%%%%%%%%%%%%%%%%%%%%%%%%%%%%%%%%%%%%%%%%%%%%%%%%%%%%%%%%%%%%%%%
%%%%%%%%%%%%%%%%%%%%%%%%%%%%%%%%%%%%%%%%%%%%%%%%%%%%%%%%%%%%%%%%%%%%%%%%%%%%%%%%

%%%%%%%%%%%%%%%%%%%%%%
\section{Conclusions}
\label{Sec-final}
%%%%%%%%%%%%%%%%%%%%%%

 Time-odd mean fields (nuclear magnetism) have been studied in rotating nuclei 
in a systematic way within the framework of CDFT. The main results can be 
summarized as follows.

\begin{itemize}

\item
  NM can considerably modify the band crossing features (crossing frequencies, 
the properties of kinematic and dynamic moments of inertia in the band 
crossing region). In the calculations without pairing, these modifications 
depend on the underlying changes in the single-particle properties such as 
alignments and energies induced by NM. These effects are 
also active in the calculations with pairing. In addition, in the calculations 
with pairing the gradual breaking of high-$j$ pairs proceeds faster in the
presence of NM, which is reflected  in a faster decrease of pairing with 
increasing $\Omega_x$. Thus we can specify this effect as {\it an anti-pairing 
effect induced by NM}.

\item
Outside the band crossing regions, the contribution of NM to the kinematic 
and dynamic moments of inertia only weakly depends on the RMF parametrization.

\item
It is shown for the first time within the self-consistent approach 
that the moments of inertia of super- and hyperdeformed configurations in 
unpaired regime come very close to the rigid-body values. Despite that the 
presence of strong vortices demonstrates the dramatic deviation of the 
currents from rigid rotation. On the contrary, the moments of inertia
of normal-deformed nuclei deviate considerably from the rigid-body value
in the calculations without pairing.

\item
Complicated structure of the currents in the rotating systems of independent 
fermions is the consequence of the fact that total current is the sum of the 
single-particle currents. The single-particle currents
show vortices (circulations), the strength and localization of which
depend on the single-particle state.

\item
  Within specific configuration the impact of NM on the binding energies
reaches its maximum at the terminating state \cite{A.08}. Underlying 
microscopic mechanism for additional binding due to NM at such states 
has the same features as those seen in low-spin one- and 
two-particle configurations of odd and odd-odd nuclei \cite{AA.10}. 
However, the magnitude of the effects is significantly larger. The 
perturbative results clearly indicate that additional binding due NM 
at terminating states is defined mainly by time-odd fields and that the 
polarization effects in fermionic and mesonic sectors of the model cancel 
each other to a large degree.

\item
 Signature-separation phenomenon in odd-odd nuclei has been analysed in detail.
It is shown that the effects neglected in the current approach such as the residual 
interaction of unpaired proton and neutron and the coupling scheme of angular momenta 
vectors of these particles at low spin considerably complicate quantitative description of the 
spectra of odd-odd nuclei. The best way to confirm the existence of this phenomenon 
would be to find (both in experiment and in calculations) the configurations of 
odd-odd nuclei which show no signature splitting in the absence of time-odd mean 
fields and measurable signature separation in the presence of time-odd mean fields.

\end{itemize}

  Although time-odd mean fields affect different physical observables (see 
introduction in Ref.\ \cite{AA.10} for details), this investigation clearly
shows that rotating nuclei still offer one of the best probes of this channel of 
density functional theories. This is because the impact of time-odd mean 
fields is significant representing on average 20\% of kinematic and dynamic
moments of inertia. In addition, it shows appreciable variations with 
configuration, particle number and rotational frequency; these variations
provide a useful tool for a better test or definition of time-odd mean fields. 
Significant amount of the data on different types (normal- \cite{AF.05}, 
superdeformed \cite{AKR.96,ALR.98,ARR.99,SZF.02,Rag.93,VALR.05}, and smooth 
terminating \cite{AFLR.99,VALR.05}) of rotational bands in unpaired regime 
available in different mass regions offers a testing ground for time-odd mean 
fields. This data is also extremely useful for fitting the parameters of time-odd 
mean fields as needed, for example, in Skyrme energy density functionals, in which 
these fields are not well defined (Refs.\ \cite{DD.95,SDMMNSS.10}). Our investigation, 
however, suggests that such fit has to be performed to a significant set of 
rotational structures representing different mass regions and different configurations 
and spanned over significant frequency range in order to minimize the dependence of 
the fit parameters on the choice of experimental data.

%%%%%%%%%%%%%%%%%%%%%%%%%%%%%% 
\section{Acknowledgments}
%%%%%%%%%%%%%%%%%%%%%%%%%%%%%%

The material is based on work supported by the Department of Energy 
under grant No. DE-FG02-07ER41459.

%%%%%%%%%%%%%%%%%%%%%%%%%%%


\begin{thebibliography}{99}
%%%%%%%%%%%%%%%%%%%%%%%%%%%

\bibitem{VALR.05} D.\ Vretenar, A.\ V.\ Afanasjev, G.\ A.\ Lalazissis, and 
P.\ Ring, Phys.\ Rep. {\bf 409}, 101 (2005).

\bibitem{BHR.03} M.\ Bender, P.-H.\ Heenen, and P.-G.\ Reinhard, Rev.\ Mod.\
Phys. {\bf 75}, 121 (2003).

\bibitem{DD.95} J.\ Dobaczewski and J.\ Dudek, Phys.\ Rev. {\bf C52}, 1827 (1995).

\bibitem{AR.00} A.\ V.\ Afanasjev and P.\ Ring, Phys.\ Rev. {\bf C62},
031302(R) (2000).

\bibitem{AA.10} A.\ V.\ Afanasjev and H.\ Abusara, Phys.\ Rev. C {\bf 81}, 014309 
                (2010).

\bibitem{KR.89} W.\ Koepf and P.\ Ring, Nucl.\ Phys. {\bf A493}, 61 (1989).

\bibitem{PWM.85} U.\ Post, E.\ W\"ust and U.\ Mosel, Nucl.\ Phys. 
{\bf A437}, 274 (1985).

\bibitem{KR.93} J.\ K{\"o}nig and P.\ Ring, Phys.\ Rev.\ Lett. {\bf 71}, 
3079 (1993).

\bibitem{AKR.96} A.\ V.\ Afanasjev, J.\ K{\"o}nig, and P.\ Ring, Nucl.\ Phys. 
{\bf A608}, 107 (1996).

\bibitem{ALR.98} A.\ V.\ Afanasjev, G.\ Lalazissis and P.\ Ring, Nucl.\ Phys. 
{\bf A634}, 395 (1998).

\bibitem{ARR.99} A.\ V.\ Afanasjev, I.\ Ragnarsson and P.\ Ring,
Phys.\ Rev. {\bf C59}, 3166 (1999).

\bibitem{AF.05} A.\ V.\ Afanasjev and S.\ Frauendorf, Phys.\ Rev.
C {\bf 71}, 064318 (2005).

\bibitem{CRHB} A.\ V.\ Afanasjev, P.\ Ring, and J.\ K\"onig, Nucl.\ Phys.
{\bf A676}, 196 (2000).

\bibitem{A.08} A.\ V.\ Afanasjev, Phys.\  Rev. {\bf C78}, 054303 (2008).

\bibitem{AA.09} H.\ Abusara and A.\ V.\ Afanasjev, Phys.\ Rev. C {\bf 79}, 024317 (2009).

\bibitem{MADLN.07} M.\ Matev, A.\ V.\ Afanasjev, J.\ Dobaczewski, G.\ A.\ Lalazissis,
and W.\ Nazarewicz, Phys.\ Rev. {\bf C 76}, 034304 (2007).

\bibitem{A250} A.\ V.\ Afanasjev, T.\ L.\ Khoo, S.\ Frauendorf, G.\ A.\ Lalazissis, 
and I.\ Ahmad, Phys.\ Rev. C {\bf 67}, 024309 (2003).

\bibitem{AA.08} A.\ V.\ Afanasjev and H.\ Abusara, Phys.\ Rev. {\bf C78}, 014315
(2008).

\bibitem{Ing.54} D.\ R.\ Inglis, Phys.\ Rev. {\bf 96}, 1059 (1954).

\bibitem{SW.86} {B.\ B.\ Serot and J.\ D.\ Walecka, Adv.\ Nucl.\ Phys.\ {\bf 16},  
1 (1986).

\bibitem{NL1} P.-G.\ Reinhard, M.\ Rufa, J.\ Maruhn, W.\ Greiner, and 
J.\ Friedrich, Z.\ Phys. A {\bf 323}, 13 (1986).

\bibitem{YM.00} M.\ Yamagami and K.\ Matsuyanagi, Nucl.\ Phys. {\bf A672}, 123 (2000).

\bibitem{CDK.08} B.\ G.\ Carlsson, J.\ Dobaczewski, and M.\ Kortelainen,
Phys.\ Rev.\ C {\bf 78}, 044326 (2008).

\bibitem{DCK.10} J.\ Dobaczewski, B.\ G.\ Carlsson and M.\ Kortelainen, 
                 J.\ Phys. {\bf G37} 075106 (2010) .

\bibitem{N.75} J.\ W.\ Negele, in {\it Effective interactions and Operators in Nuclei},
Lecture Notes in Physics 40 (Springer, Berlin, 1975), p.250

\bibitem{AFLR.99} A.\ V.\ Afanasjev, D.\ B.\ Fossan, G.\ J.\ Lane and  I.\ Ragnarsson, 
Phys.\ Rep. 322, 1 (1999).

\bibitem{SW.05} W.\ Satula and R.\ Wyss, Rep.\ Prog.\ Phys. {\bf 68}, 131 (2005).

%%%%%%%% D1S set of Gogny force
\bibitem{D1S} J.\ F.\ Berger, M.\ Girod, and D.\ Gogny,
Comp.\ Phys.\ Comm. {\bf 63}, 365 (1991).

\bibitem{SDMMNSS.10} N.\ Schunck, J.\ Dobaczewski, J.\ McDonnell, J.\ M\'{o}re, W.\ Nazarewicz,
J.\ Sarich, and M.\ V.\ Stoitsov, Phys.\ Rev. {\bf C 81}, 024316 (2010).

\bibitem{DDW.03} J.\ Dobaczewski, J.\ Dudek, and R.\ Wyss, Phys.\ Rev. C {\bf 67}, 
034308 (2003).

\bibitem{Pingst-A30-60} A.\ V.\ Afanasjev, P.\ Ring and I.\ Ragnarsson,
Proc. Int. Workshop PINGST2000 "Selected topics on $N=Z$ nuclei", 
2000, Lund, Sweden, Eds. D.\ Rudolph and M. Hellstr{\"o}m, (2000) p.\ 183.

\bibitem{DFPCU.04} M.\ A.\ Delaplanque, S.\ Frauendorf, V.\ V.\ Pashkevich, S.\ Y.\ Chu,
and A.\ Unzhakova, Phys.\ Rev. C {\bf 69}, 044309 (2004).

\bibitem{NL3*} G. \ A. \ Lalazissis, S. \ Karatzikos, R. \ Fossion, D. \ Pena  Arteaga,  
A. \ V. \ Afanasjev, and P. \ Ring,  Phys. \ Lett. {\bf B671}, 36 (2009).

\bibitem{AKRRE.00} A.\ V.\ Afanasjev,  J.\ K{\"o}nig, P.\ Ring, L.\ M.\ Robledo,
J.\ L.\ Egido, Phys.\ Rev.\ C {\bf 62}, 054306 (2000).

\bibitem{Rag.93} I.\ Ragnarsson, Nucl.\ Phys. {\bf A557}, c167 (1993).

\bibitem{BM.book} A.\ Bohr and B.\ Mottelson, Nuclear Structure, vol.\ II, Benjamin,
New York (1975).


\bibitem{KG.77} K.-K.\ Kan and J.\ J.\ Griffin, Phys.\ Rev.\ C {\bf 15}, 
1126 (1977).

\bibitem{DSK.85} M.\ Durand, P.\ Schuck, and J.\ Kunz, Nucl.\ Phys. {\bf A439}, 
                 263 (1985).

\bibitem{MQS.97} I.\ N.\ Mikhailov, P.\ Quentin, and D.\ Samsoen, Nucl.\ Phys.\
{\bf A627}, 259 (1997).

\bibitem{LSQM.03} H.\ Laftchiev, D.\ Samsoen, P.\ Quentin, and I.\ N.\ Mikhailov,
Phys.\ Rev. C {\bf 67}, 014307 (2003).

\bibitem{R.76} M.\ Radomski, Phys.\ Rev. {\bf 14}, 1704 (1976).

\bibitem{GR.78-0} P.\ Gulshani and D.\ J.\ Rowe, Can.\ J.\ Phys. {\bf 56}, 468 (1978).

\bibitem{GR.78} P.\ Gulshani and D.\ J.\ Rowe, Can.\ J.\ Phys. {\bf 56}, 480 (1978).

\bibitem{KM.79} J.\ Kunz and U.\ Mosel, Nucl.\ Phys. {\bf A323}, 271 (1979).

\bibitem{FKMW.80} J.\ Fleckner, J.\ Kunz, U.\ Mosel, and E.\ Wuest, Nucl.\ Phys. 
{\bf A339}, 227 (1980).

\bibitem{Zn60}  C.\ E.\ Svensson {\it et al},
Phys.\ Rev.\  Lett. {\bf 82}, 3400 (1999).

\bibitem{NL3} G.\ A.\ Lalazissis, J.\ K\"onig and P.\ Ring, Phys.\ Rev. C {\bf 55}, 
540 (1997).

\bibitem{NLSH} M.\ M.\ Sharma, M.\ A.\ Nagarajan and P.\ Ring,  Phys.\ Lett. {\bf B312}, 
377 (1993).

\bibitem{NLZ} M.\ Rufa, P.-G.\ Reinhard, J.\ A.\ Maruhn, W.\ Greiner, and M.\ R.\ Strayer, 
Phys.\ Rev. {\bf C38}, 390 (1988).

\bibitem{Dy154} Q.\ A.\ Ijaz} {\it et al}, Phys.\ Rev. C {\bf 80}, 034322 (2009).

\bibitem{CHO} T.\ Troudet and R.\ Arvieu, Annals of Physics {\bf 134}, 1 
(1981).

\bibitem{Ragbook} S.\ G.\ Nilsson and I.\ Ragnarsson, {\it Shapes 
and Shells in Nuclear Structure}, Cambridge University Press, 1995.

\bibitem{Kr74-no-term} J.\ J.\ Valiente-Dob\'on {\it et al}, Phys.\  Rev.\ Lett. 
                       {\bf 95} 232501 (2005).

\bibitem{S.book} Z.\ Szymanski, ``Fast Nuclear Rotations'', Oxford studies in
physics, Claredon press, Oxford, 1983.

\bibitem{And78}  C.\ G.\ Andersson, G.\ Hellstr\"{o}m, G.\ Leander, I.\ Ragnarsson, 
S.\ \AA berg, J.\ Krumlinde, S.\ G.\ Nilsson and Z.\ Szyma\'{n}ski, Nucl.\ Phys. {\bf A309}, 
             141 (1978).

\bibitem{RS.80}
P.\ Ring and P.\ Schuck, {\it The Nuclear Many-body Problem\/}, Springer-Verlag,
Heidelberg (1980).

\bibitem{VDS.83}  M.\ J.\ A.\ de Voigt, J.\ Dudek and Z.\ Szyma\'{n}ski, Rev.\ Mod.\ Phys.  
                 {\bf 55}, 949 (1983).

\bibitem{ZSW.05}  H.\ Zdunczuk, W.\ Satula, and R.\ A.\ Wyss, Phys.\ Rev. {\bf C71}, 
                  024305 (2005).

\bibitem{RGL.97} P.\ Ring, Y.\ K.\ Gambhir, and G.\ A.\ Lalazissis, 
Comp.\ Phys.\ Comm. {\bf 105}, 77 (1997).

\bibitem{MDD.00} H.\ Molique, J.\ Dobaczewski and J.\ Dudek, Phys.\ Rev. {\bf C 61}, 
                 044304 (2000).

\bibitem{SDNR.10} W.\ Satula, J.\ Dobaczewski, W.\ Nazarewicz, and M. Rafalski,
Phys.\ Rev. C {\bf 81}, 054310 (2010).

%\bibitem{S.98} C.\ E.\ Svensson {\it et al}, Phys.\ Rev. C {\bf 58}, R2621 (1998).

%\bibitem{L.96} S.\ M.\ Lensi {\it et al}, Z.\ Phys. {\bf A354}, 117 (1996).

%\bibitem{CEMPRR.95} E.\ Caurier, J.\ L.\ Egido, G.\ Martinez-Pinedo, A.\ Poves, J.\ Retamosa,
%L.\ M.\ Robledo, and A.\ P.\ Zuker, Phys.\ Rev.\ Lett. {\bf 75}, 2466 (1995).

%\bibitem{L.99} S.\ M.\ Lensi {\it et al}, Phys.\ Rev. C {\bf 60}, 021303 (1999). 

\bibitem{GM.58} C.\ J.\ Gallagher Jr., and S.\ A.\ Moszkowski, Phys.\ Rev. 
{\bf 111}, 1282 (1958).

% Coriolis coupling in two-quasiparticle rotational bands of deformed even-even
% nuclei
\bibitem{GJ.92} A.\ Goel and A.\ K.\ Jain, Phys.\ Rev. {\bf C 45}, 221 (1992).

\bibitem{BPO.76} J.\ P.\ Boisson, R.\ Piepenbring and W.\ Ogle, Phys.\ Rep.
{\bf 26}, 99 (1976).

\bibitem{NKSSN.94} D.\ Nosek, J.\ Kvasil, R.\ K.\ Sheline, P.\ C.\ Sood, and
J.\ Noskov{\'a}, Int.\ J.\ Mod.\ Phys.\ {\bf E3}, 967, 1994.

\bibitem{N.62} N.\ D.\ Newby Jr., Phys.\ Rev. {\bf 125}, 2036 (1962).

% Nuclear structure in odd-odd nuclei, 144<A<194
\bibitem{JSHSet.98} A.\ K.\ Jain, R.\ K.\ Sheline, D.\ M.\ Headly,  P.\ C.\ Sood, 
D.\ G.\ Burke, I.\ Hr{\v{i}}vn{\'a}cov{\'a}, J.\ Kvasil, and D.\ Nosek,
R.\ W.\ Hoff, Rev.\ Mod.\ Phys. {\bf 70}, 843 (1998).

% Coriolis coupling in the rotational bands of deformed odd-odd nuclei
\bibitem{JKSH.89} A.\ K.\ Jain, J.\ Kvasil, R.\ K.\ Sheline, and R.\ W.\ Hoff,
Phys.\ Rev. {\bf C 40}, 432 (1989).  

\bibitem{BLMQ.87} L.\ Bennour, J.\ Libert, M.\ Meyer, and P.\ Quentin, Nucl.\ Phys.
                  {\bf A465}, 35 (1987).

\bibitem{F.01} S.\ Frauendorf, Rev.\ Mod.\ Phys. {\bf 73}, 463 (2001).

\bibitem{SZF.02} B.\ Singh, R.\ Zywina and R.\ B.\ Firestone, Nucl.\ Data Sheets 
                 {\bf 97}, 241 (2002).

\end{thebibliography}
\end{document}